\documentclass[sigplan, nonacm]{acmart}
\AtBeginDocument{%
  }

\usepackage[most]{tcolorbox}
\usepackage{dsfont}
\usepackage{amsmath,amssymb,amsfonts}
\usepackage{algorithmic}
\usepackage{graphicx}
\usepackage{textcomp}
\usepackage{xcolor}
\usepackage{circledsteps}
\usepackage{subfigure}
\usepackage{xspace}

\newcounter{question}

\title[]{Designing Datacenter Power Delivery \\ Hierarchies for the AI Era}

\author{Grant Wilkins}
\authornote{Work partly completed while an intern at Microsoft Azure Research.}
\affiliation{%
  \institution{Stanford University}
  \city{Stanford}
  \state{CA}
  \country{USA}
}

\author{Fiodar Kazhamiaka}
\affiliation{%
  \institution{Microsoft Azure Research}
  \city{Redmond}
  \state{WA}
  \country{USA}
}

\author{Alok Gautam Kumbhare}
\affiliation{%
  \institution{Microsoft Azure Research}
  \city{Redmond}
  \state{WA}
  \country{USA}
}

\author{Chaojie Zhang}
\affiliation{%
  \institution{Microsoft Azure Research}
  \city{Redmond}
  \state{WA}
  \country{USA}
}

\author{Ricardo Bianchini}
\affiliation{%
  \institution{Microsoft Azure Research}
  \city{Redmond}
  \state{WA}
  \country{USA}
}

\newcommand{\azure}[0]{Azure\xspace}

\begin{abstract}
Demand for AI accelerators is rapidly increasing rack power density, with 
projections approaching 1MW per deployment by 2027. This poses a major
challenge for datacenter power delivery designers.
As power densities increase, a datacenter designed for a different target density
may strand power, i.e., may be unable to use all the power that its delivery 
hierarchy has provisioned. Designs must remain efficient over long datacenter 
lifetimes and multiple hardware generations. Power utilization is particularly important as grid 
power capacity is a scarce resource in the AI era.

Designing an efficient power delivery hierarchy for the long run is difficult 
because rack placement feasibility, workload impact, and cost depend jointly on 
electrical topology, deployment granularity, placement policy, power 
oversubscription, and workload mix. Moreover, each of these factors evolve over time, 
have inter-dependencies across multiple resource dimensions, and generally do not lend themselves to closed-form 
analysis.

To address this challenge, we develop a framework for evaluating datacenter power delivery designs using throughput, power, and cost metrics over realistic arrival, oversubscription, and decommissioning sequences. The framework combines projection models for GPU, compute, and storage deployments with operational factors grounded in production data from \azure. Our results show that multi-resource stranding materially changes deployable capacity, effective capital expenditure, and delivered performance, and quantify how rising density from rack- and pod-scale AI systems shapes these outcomes. For AI datacenter design, the relevant planning objective is not installed megawatts, but deployable capacity over time.

\end{abstract}

\begin{document}

\maketitle

\section{Introduction}

\begin{figure}
    \centering
    \includegraphics[width=0.95\columnwidth]{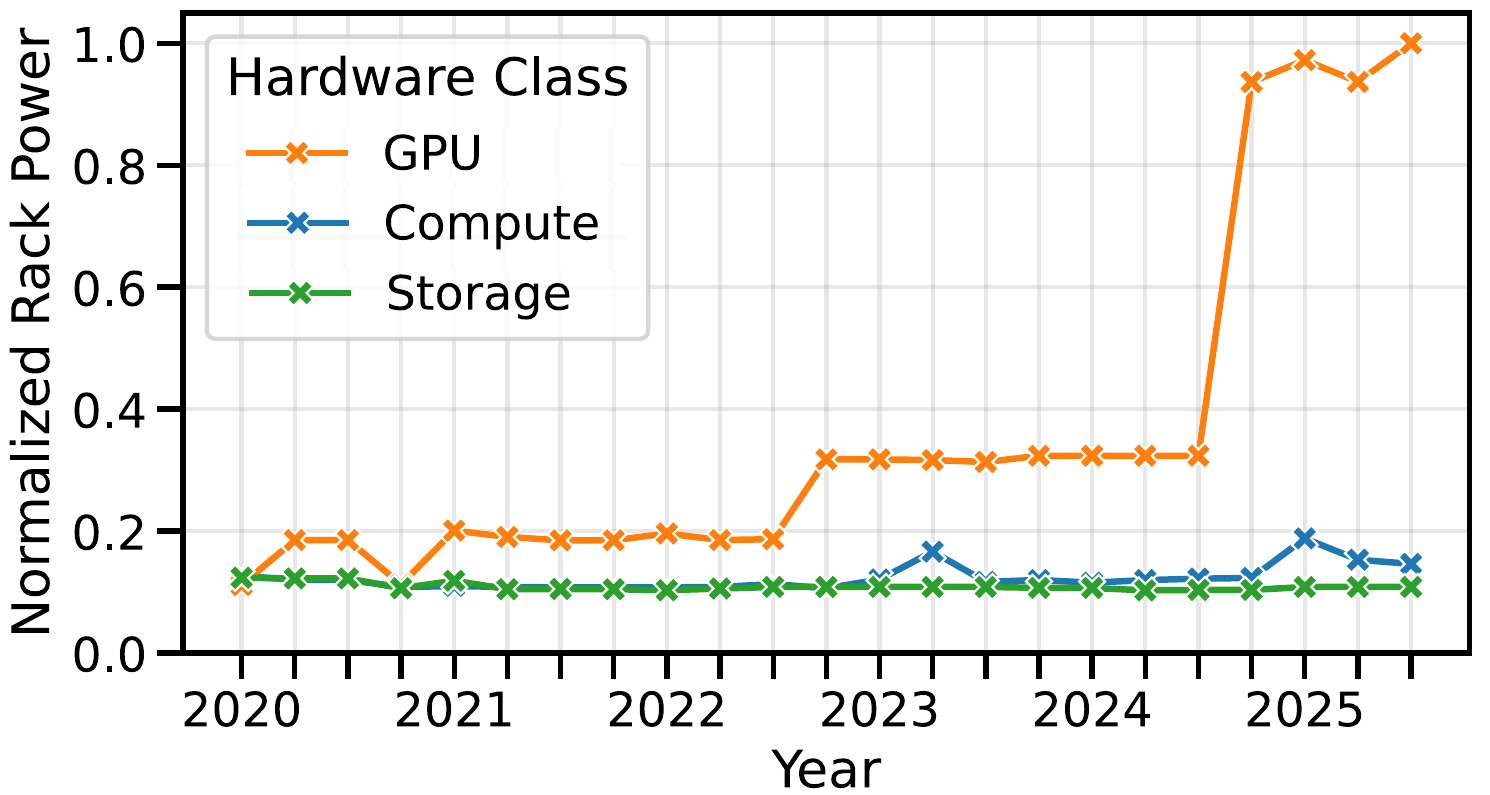}
    \caption{P99 of rack power density since 2020 for datacenter deployments, showing distinct accelerator generations and a widening gap between GPU and non-GPU power density. Density is normalized to the maximum P99 value observed in each quarter at \azure.}
    \label{fig:power-density-p99}
\end{figure}

\textbf{Motivation.}
Demand for generative AI is driving a rapid datacenter buildout. These new AI 
datacenters are substantially different than their cloud datacenter predecessors 
with respect to their power delivery hierarchies.
In the web-services era, rack power density was often below 20\,kW~\cite{barroso2019datacenter}. Today, AI accelerator racks exceed 150\,kW, and public roadmaps project rack- and pod-scale systems approaching 1\,MW in a few years~\cite{nvidia-rubin-ultra, nvidia-rubin, semianalysisrubin}. Figure~\ref{fig:power-density-p99} shows the same shift in production deployments at \azure, where accelerator rack power is rising much faster than general compute and storage racks. These denser systems also tighten infrastructure coupling, as power, cooling, and local interconnect increasingly scale together~\cite{nvidia-800vdc,ocp-summit-2025,ocp2023oailiquidcooling}.

Datacenter halls are built with a fixed power delivery hierarchy that is chosen at construction time and persists for 15--25 years across multiple hardware generations~\cite{turnerandtownsend2024index, CushWakeCostGuide_2025}. 
Power delivery design is thus a long-term commitment. A hierarchy chosen for today’s hardware must remain efficient for substantially denser systems that arrive years later.


Worse, rack deployment feasibility is hierarchical, not site- or even hall-wide. A rack or pod must satisfy capacity and redundancy constraints at every level of the power delivery path~\cite{tier,zhang2021flex,baxi2025onlinerackplacementlargescale}. As deployment quanta grow to consume a meaningful fraction of UPS, busbar, PDU, and cooling capacity, aggregate provisioned power becomes a misleading proxy for what the hall can still admit. A hall can retain substantial available power and still reject the next deployment because the remaining capacity is fragmented across domains. We  refer to this unusable capacity as \textit{stranded}.

Standard planning metrics do not capture this effect. Provisioned MW and CapEx per MW price installed electrical capacity, not the load a hierarchy can continue to admit after years of arrivals and partial filling~\cite{barroso2019datacenter}. A hall can therefore look efficient at commissioning and still perform poorly later, if future hardware cannot use the residual capacity structure left by earlier placements. This becomes harder to ignore as deployment quanta grow significantly and fast. The problem is also dynamic: hall efficiency depends on the sequence of hardware arrivals and retirements, power oversubscription (harvesting), and prior placements across electrical domains~\cite{baxi2025onlinerackplacementlargescale}. These interactions form a multi-resource, multi-year packing problem over power, liquid cooling, air cooling, and space, and does not admit a closed-form analysis.

\begin{figure}
    \centering
    \includegraphics[width=\columnwidth]{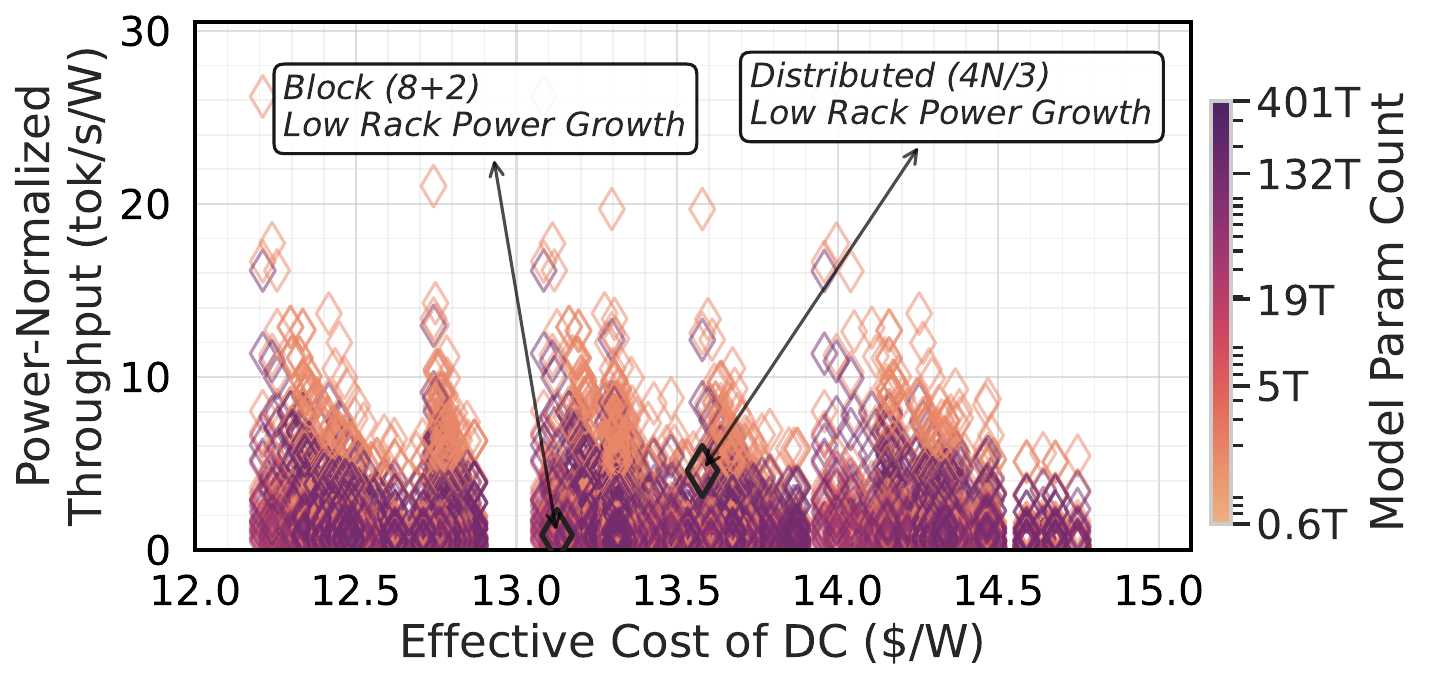}
    \caption{Each marker represents a combination of datacenter design, workload, and rack density projections. Colors represent different LLM mixture of experts models being served across the fleet. Each combination is analyzed with our framework, and is compared on throughput of LLM inference per watt versus effective fleet cost. Highlighted points represent diverse workloads, labels describe the points' design and power projection.}
    \label{fig:space-complexity}
\end{figure}

Workload throughput metrics (such as inference tokens per second), which ultimately capture the purpose of datacenters, add another dimension to the design problem.
Throughput depends on the deployed hardware, and power delivery designs that do not support high power density may be cheaper to provision on a per‑MW basis. 
This introduces a design trade-off in the tokens/sec/W metric as shown in Figure~\ref{fig:space-complexity}: the higher throughput enabled by hosting workloads on a high-power, tightly interconnected pod must be balanced against the increased infrastructure cost required to support such deployments. 
The size, architecture, and parallelism configurations of the models being hosted will influence this trade-off.

{\bf Our work.} In light of these gaps and challenges, we propose a framework for power delivery
design and evaluation that accounts for all relevant factors to produce the most
efficient design over the datacenter's lifetime.  We model the designs by the capacity they can actually \textit{deploy} over time, not just the capacity they \textit{install} at commissioning. Our framework simulates multi-year fleet evolution under hierarchical multi-resource placement constraints and combines infrastructure cost models with comparative throughput models for emerging accelerator generations. 

Using our framework, we evaluate many variants of two common classes of power delivery designs: distributed redundant and block redundant. Our results show that designs with similar provisioned capacity can diverge significantly in deployable capacity, effective cost, and delivered throughput. Figure~\ref{fig:space-complexity} shows how important metrics such as throughput per watt and cost per watt can differ across a variety of power delivery designs, rack power projections, and LLM inference workloads: throughput per watt varies by more than 20x while cost per watt varies by more than 20\%.  The impact of these ranges is measured in billions of dollars. 

In summary, we make the following contributions:
\begin{itemize}
    \item We identify \emph{deployable power capacity over time}, rather than installed MW or first-order \$/W, as the key metric for evaluating AI datacenter power delivery designs.
    \item We develop an evaluation framework that quantifies this metric under hierarchical multi-resource placement constraints with arrivals, oversubscription, and decommissioning.
    \item We show that redundancy topology changes how capacity becomes stranded as deployment quanta grow: distributed-redundant designs degrade through fragmented headroom, and block-redundant through coarse capacity granularity.
    \item We quantify when larger GPU pods improve inference efficiency enough to justify their higher deployment granularity under realistic power-delivery constraints.
\end{itemize}

\section{Background}

Modern datacenters are built to satisfy two requirements at once: high availability under equipment or utility failures, and sufficient distribution capacity to host rack deployments. Those requirements are related but not identical. A design may provision substantial aggregate power and still be unable to admit a new deployment due to reserve, row, or line-up constraints. This paper studies how that gap emerges over time as deployments are placed into a fixed hierarchy.

\subsection{Power-Delivery Hierarchy and Feasibility}
\begin{figure}
    \centering
    \includegraphics[width=\columnwidth]{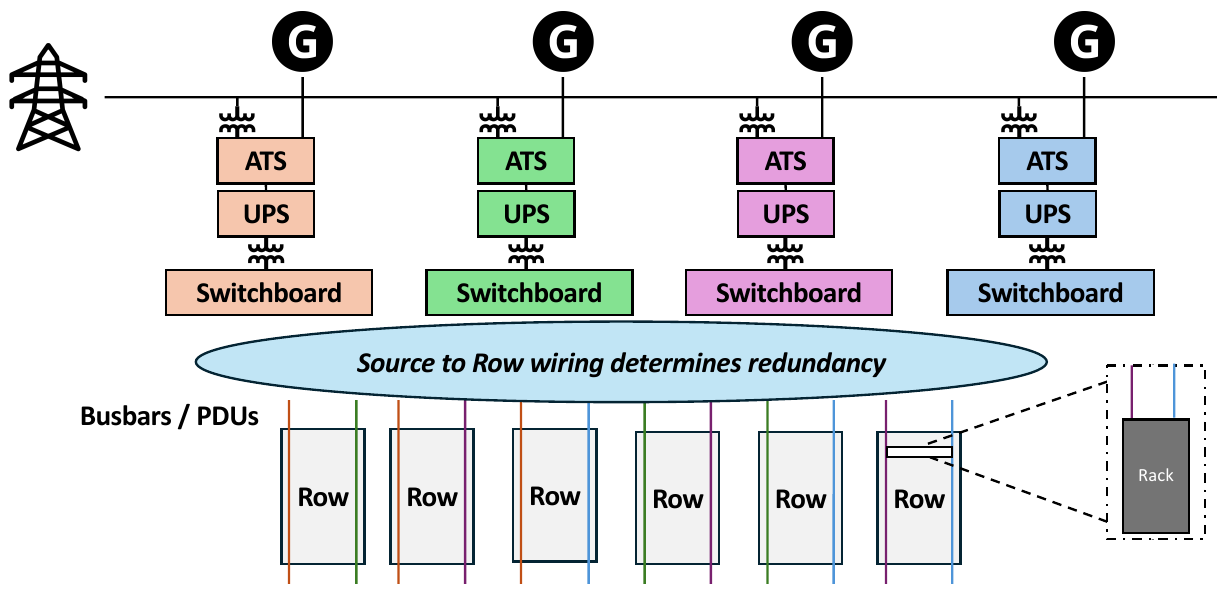}
    \caption{Example of major components in a datacenter power-delivery hierarchy, from grid and generator/battery down to the rack level. Generic diagram not showing redundancy.}
    \label{fig:electrical-hierarchy}
\end{figure}

Datacenters are organized as one or more data halls served by a power-delivery hierarchy with some components shared across halls, (e.g., generators or batteries). Designers provision redundancy throughout this hierarchy so that failures, maintenance, or upstream outages do not interrupt service. We do not model outage probability directly. Instead, we study how standard Tier III/IV-style availability choices shape effective deployable capacity~\cite{tier}. A typical power path, shown in Figure~\ref{fig:electrical-hierarchy}, includes stages such as Grid $\rightarrow$ Substation $\rightarrow$ Backup Generation/Storage $\rightarrow$ UPS $\rightarrow$ Switchboard $\rightarrow$ Row Distribution $\rightarrow$ Rack PSU $\rightarrow$ Server PSU, though the exact ordering varies across facilities~\cite{polca, zhang2021flex, dynamo2016wu, li2019capmaestro, google-mv, fan2007power,barroso2019datacenter}. We use \textit{line-up} to refer to a common upstream electrical branch: a set of rows or racks that share the same upstream power-delivery equipment. Deployment feasibility is hierarchical, as a deployment must satisfy capacity and redundancy constraints at every level of the path, not only at the site interconnect.

Racks are typically deployed in groups referred to as \emph{clusters} that are in close proximity, ideally within a single row and connected to the same row-level network switch.
This deployment pattern emphasizes row power capacity as a key constraint, where power capacity can become stranded in row-level pockets.

\subsection{Supporting High-Power Racks}
Typically, each row of racks is supplied by two busbars connected to independent line-ups. At 400~V AC distribution, individual busbar capacity is typically limited to less than 1~MW, owing to safety, installation, and supply-chain constraints. Supporting higher-power deployments within a single row therefore requires installing additional busbars in parallel.
Notably, this approach can fragment capacity and reduce placement flexibility: high-power accelerators are preferentially deployed in costly high-density (HD) rows provisioned with greater power capacity, while lower-power CPU and storage servers are placed in low-density (LD) rows.

\subsection{Redundancy as a Capacity Constraint}
\begin{figure}
    \centering
    \subfigure[Distributed Redundant (4$N$/3)]
    {\includegraphics[width=0.7\linewidth]{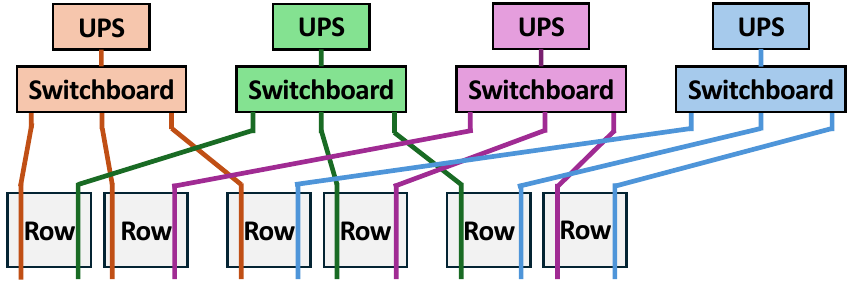}\label{fig:dist-red}}
    \subfigure[Block Redundant (3+1)]
    {\includegraphics[width=0.7\linewidth]{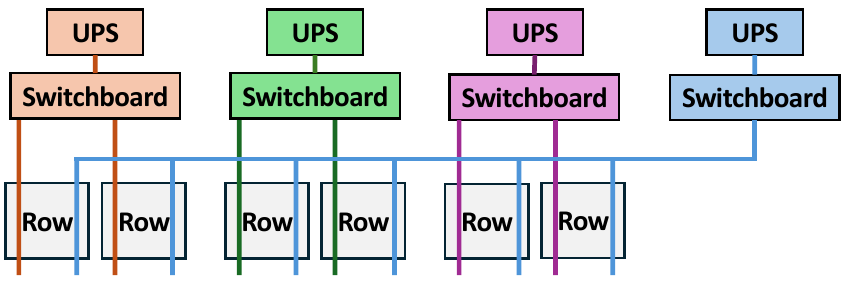}\label{fig:block-red}}
    \caption{Schematic wiring differences between (a) distributed and (b) block redundant designs. In distributed designs, failover is managed by keeping reserve load on each UPS. In block designs, failover is managed by transferring load to the reserve UPS.}
    \label{fig:placeholder}
\end{figure}

Power delivery hierarchies commonly use either \emph{distributed redundancy} or \emph{block redundancy}~\cite{barroso2019datacenter, polca, zhang2021flex} to provide high availability. The relevant distinction, as rack power increases, is that the design affects how much residual headroom remains usable for new placements. In \emph{distributed redundancy}, reserve is spread across active line-ups. An $xN/y$ design provides $x$ total line-ups but only $y$ line-ups of supported load, so each parent must retain some capacity for fail-overs. For example, in a $4N/3$ system, as shown in Figure~\ref{fig:dist-red}, each line-up reserves $25\%$ of its capacity so the system can survive the loss of any one line-up~\cite{zhang2021flex}. This makes reserve flexible, but it reduces total usable capacity within each line-up. In \emph{block redundancy} (Figure~\ref{fig:block-red}), some line-ups are used only for fail-overs. This avoids sharing reserve across active domains, and increases the total capacity in each line-up, but makes capacity coarser and less flexible to use across blocks~\cite{barroso2019datacenter, polca}. The tradeoff is therefore structural: distributed designs tend to fragment residual headroom across active domains, while block designs tend to quantize usable capacity at coarser units.

\subsection{Capacity Stranding}

\emph{Stranded capacity} is provisioned capacity that remains unused because another constraint binds first~\cite{li2019capmaestro}. In datacenters, stranding is multi-level: a facility may retain headroom at the site or line-up level and still be unable to admit another rack because some lower-level constraint binds first. This matters because modern AI deployments are coarse placement units. They arrive as racks or pods, often with joint networking and cooling requirements~\cite{nvidia-800vdc,nvidia-rubin,nvidia-rubin-ultra}. As a result, deployability depends not only on how much infrastructure is installed, but also on how that capacity is partitioned across the hierarchy.



\section{How Power Delivery Topologies Strand Capacity}
\label{sec:stranding-mechanisms}

Datacenter electrical designs are often compared by installed high-availability (HA) capacity and CapEx per provisioned megawatt. We show why those commissioning metrics can be misleading, isolate and describe topology-specific mechanisms that lead to deployment inefficiencies, and motivate the lifecycle evaluation framework that follows.

\subsection{A Tale of Two Designs}
\label{sec:static-mislead}

\begin{figure}
\subfigure[Single Data Hall]{
\label{fig:steady-q0}{\includegraphics[width=0.48\columnwidth]{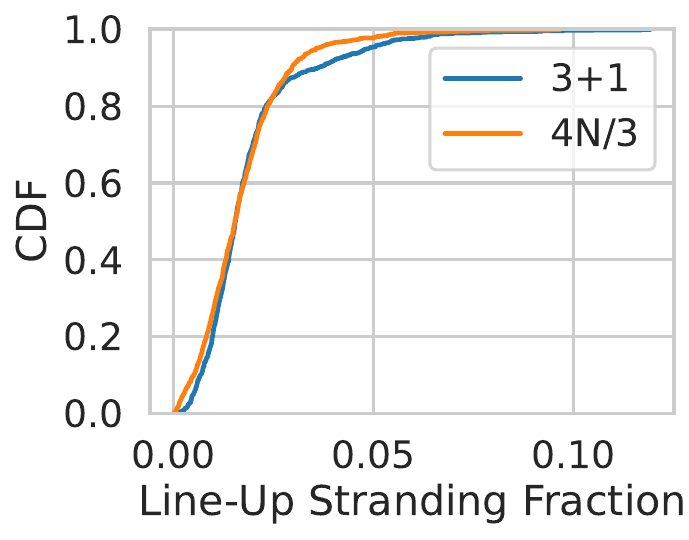}}}
\subfigure[Fleet-Wide, Lifecycle]{
\label{fig:strand-fleet}{\includegraphics[width=0.48\columnwidth]{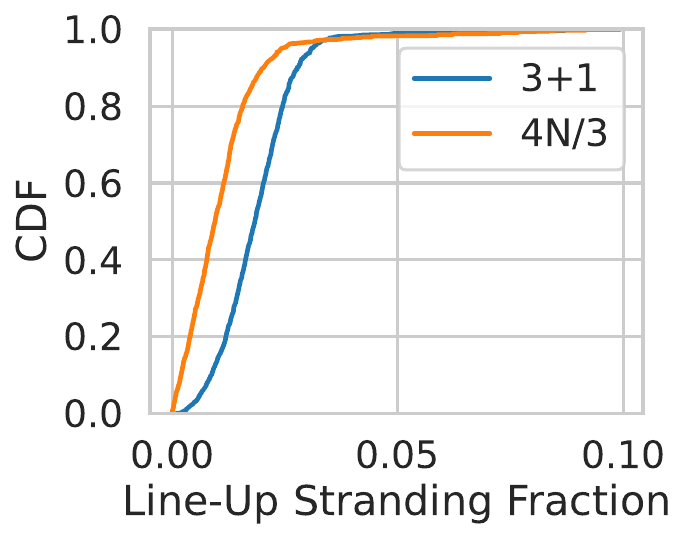}}}
\caption{CDF of UPS stranding under (a) single-hall Monte Carlo analysis and (b) the final state of an 8-year fleet-scale lifecycle simulation. The local view suggests that $4N/3$ and 3+1 are similar. The lifecycle simulation separates them: 3+1 develops higher tail stranding and requires additional halls to serve the same deployed demand.}
\label{fig:question0}
\end{figure}

Consider a $4N/3$ distributed-redundant hall (as shown in Figure~\ref{fig:dist-red}) and a $3+1$ block-redundant hall (as shown in Figure~\ref{fig:block-red}). At 2.5~MW per UPS line-up, both designs provide $7.5$~MW of high-availability IT capacity and have similar baseline cost under our component model (\$10M/MW for $4N/3$ and \$10.3M/MW for 3+1); on these static metrics, $4N/3$ appears slightly preferable. A similar conclusion is reached when conducting a single-hall Monte Carlo analysis of rack placement: deployment traces of rack clusters can be generated based on projected distributions of power, cooling, and space requirements and placed into the hall until it saturates. An example of this analysis is shown in Figure~\ref{fig:steady-q0}, where the two designs exhibit similar line-up stranding, with $4N/3$ appearing slightly better.

The comparison changes once the same designs are evaluated over an 8-year fleet lifecycle. The lifecycle setting adds the effects that commissioning metrics omit: halls partially fill, hardware generations arrive with different power densities, racks are harvested or decommissioned, and new deployments must fit into the residual capacity left by earlier placements. In Figure~\ref{fig:strand-fleet}, 3+1 develops materially higher tail stranding and requires more halls to serve the same demand.

Static metrics suggest only a small difference: 3+1 costs about 3\% more per provisioned MW than $4N/3$. Over the 8-year fleet lifecycle, however, that small first-order gap nearly doubles to a 5.8\% difference in CapEx, as higher stranding in 3+1 requires 23 more halls to be built. The reason is not aggregate unused power, but placement feasibility within a partially filled hierarchy. As deployments arrive over time, admission depends on where residual headroom remains across specific UPS domains, rows, and line-ups, as restricted by the topology's failover requirements.
The mechanisms that affect stranded power are different across distributed and block designs, and are discussed next.


\subsection{Distributed Designs: Reserve Fragmentation}
\label{sec:reserve-distributed}

In a distributed $xN/y$ design, reserve is shared across active line-ups. Each line-up can only use a fraction of its rating for HA load, and the remaining must be available for failover. This makes reserve flexible, but it also makes placement require satisfying many simultaneous headroom constraints.

Consider a deployment $r$ connected to $k_r \ge 2$ parents. If one parent fails, the surviving parents must absorb the deployment's failover load. The required headroom per surviving parent is
\begin{equation}
\Delta(P_r,k_r) = \frac{P_r}{k_r-1}.
\label{eq:failover_delta}
\end{equation}

Placement is feasible only if enough parents simultaneously have at least this much local headroom. Aggregate slack is not sufficient. A hall can have $\sum_d h_d > P_r$ and still reject the deployment because the slack is spread across parents that are each individually too full.
For example, a $10N/8$ hall with ten 2.5\,MW UPS units provides 20\,MW of HA capacity. At 18\,MW deployed uniformly, each UPS has 200\,kW of headroom. A 650\,kW rack with $k_r=4$ needs about 217\,kW of headroom on each surviving parent, so placement fails despite 2\,MW of aggregate remaining capacity.

This mechanism becomes more important as deployment quanta grow. Larger racks and pods increase $\Delta(P_r,k_r)$, while partially filled halls reduce the local headroom available at each parent. Load-balancing heuristics can delay this failure by keeping headroom even across parents, but they cannot remove it once an indivisible deployment is large relative to the remaining local capacity.

\subsection{Block Designs: Line-Up Quantization}
\label{sec:block-fragmentation}

Block-redundant designs fail differently. Reserve capacity is separated from active load, and block designs avoid the cross-parent reserve-fragmentation condition above. The downside to block designs is that usable capacity is coarser. A 2 MW deployment must fit inside the remaining capacity of a single active line-up, whereas in a distributed-redundant design this power can be split across two or more line-ups.
Specifically, for a block with usable capacity $C$ and deployment power $P_r$, the block admits $\lfloor C/P_r \rfloor$ deployments. The leftover capacity is
\begin{equation}
\eta(P_r) = \left(C - \lfloor C/P_r \rfloor P_r \right)/ C.
\label{eq:eta_sawtooth}
\end{equation}

The key point is divisibility. Just below a threshold $P_r = C/q$, $q$ deployments fit. Just above the threshold, only $q-1$ fit, and the remainder of provisioned line-up capacity is insufficient for the next same-sized deployment.

\subsection{Mechanisms in Isolation}
\label{sec:mechanism-isolation}

\begin{figure}[t]
    \includegraphics[width=\columnwidth]{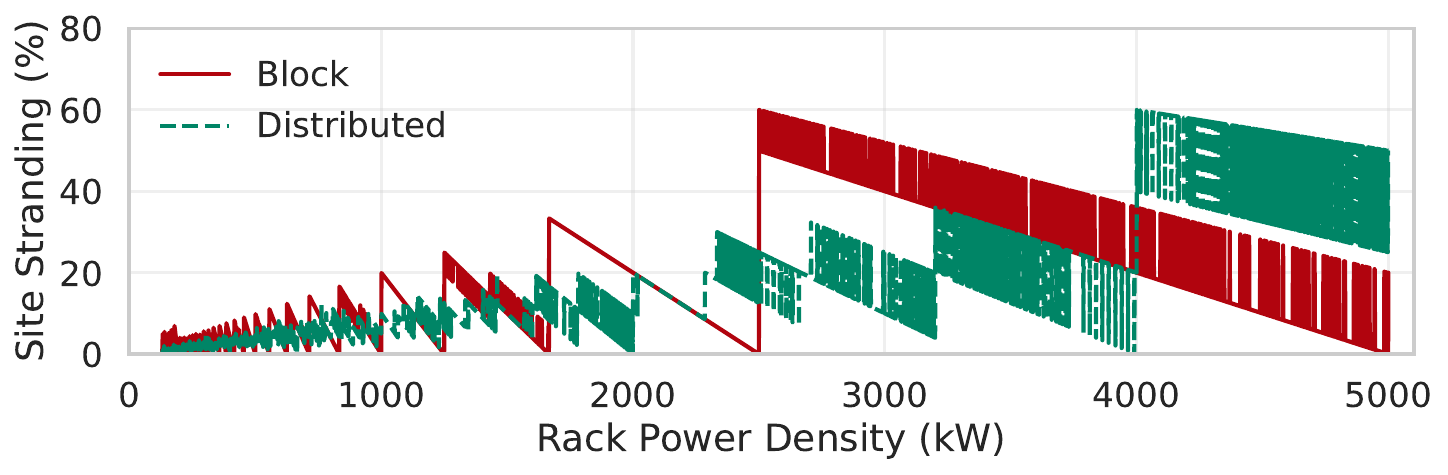}
    \caption{Single-hall, single-SKU stranding under increasing deployment power. Each experiment fills one hall with repeated deployments of the same SKU and reports the capacity left undeployable at saturation. Distributed redundancy (4$N$/3) strands capacity when too few parents have enough simultaneous failover headroom. Block redundancy (3+1) strands capacity at divisibility thresholds of the line-up or UPS-block capacity. The diagonal oscillations are due to row-space constraints.}
    \label{fig:harmonic-stranding}
\end{figure}

Figure~\ref{fig:harmonic-stranding} exposes these mechanisms with a single-SKU sweep. For each point on the x-axis, we instantiate a hall of a fixed topology and repeatedly place identical deployments of that power until placement fails. We then measure the fraction of provisioned capacity that remains unused but cannot admit another deployment. The block design shows sharp jumps because small changes in deployment power can cross a divisibility threshold. The distributed design changes more smoothly because failures arise from simultaneous headroom constraints across parents rather than a single block boundary.

This experiment is intentionally simpler than a real fleet. It removes mixed SKUs, arrival order, harvesting, retirement, and workload performance. That simplification is useful because it separates the two structural causes of stranding. In distributed redundancy, stranding rises when no sufficiently large set of parents has enough simultaneous failover headroom. In block redundancy, stranding spikes when deployment power crosses a divisibility threshold of the usable block capacity. The same symptom---undeployable power---therefore comes from different structural causes.

Real fleets smooth these patterns, but cannot undo them at high power densities. CPU and storage racks can absorb some residual fragments, and heterogeneous GPU generations reduce the sharpness of any single threshold. At the same time, online arrivals, indivisible deployment quanta, harvesting, and decommissioning make the residual capacity of each hall history-dependent. A design therefore cannot be evaluated only by installed MW or by a single saturated-hall experiment; it must be evaluated by how much capacity remains deployable after a sequence of placements and removals. The rest of the paper builds this lifecycle evaluation.

\section{Datacenter Design Evaluation Framework}\label{sec:framework}

\label{sec:foundations}
We now define the lifecycle model used to evaluate deployable capacity under hierarchical multi-resource placement constraints. The model captures three IT classes---GPU, general compute, and storage---and abstracts the arrival, placement, harvesting, and retirement processes that determine how halls fill over time. Input models for arrivals, cost, and workload impact are described in Section~\ref{sec:implementation}.

\subsection{Rack Resource and Lifetime Model}
\label{sec:arrivals}

\textbf{Power and cooling.}
Deployments are modelled at rack granularity.\footnote{For brevity, we use racks as the basic deployment unit. GPU pods are composed of multiple networked racks that must be placed together.} Each rack $r$ is characterized at installation time $\tau$ by power demand $P_r(\tau)$ and a cooling demand vector $C_r(\tau)$ derived from $P_r(\tau)$. Cooling demand is split into air cooling for storage, general compute, and GPU networking, and direct-to-chip liquid cooling for GPU accelerators. Cooling demand is calculated from rack power using fixed conversions: 165 CFM/kW for air cooling and 2 LPM per rack for direct-to-chip liquid cooling~\cite{ocp2023oailiquidcooling}. Tile space and networking ports are modeled similarly.

\textbf{Availability and connections.}
Racks are deployed at one of two availability tiers. \emph{High-availability} (HA) racks are admitted only when placement preserves the reserve required to survive any single line-up failure. \emph{Low-availability} (LA) racks may consume reserve capacity, as in Flex~\cite{zhang2021flex}, but are not guaranteed power through failures or maintenance. GPU pods require multiple independent feeds and sufficient busbar capacity, so they are restricted to high-density rows. Post-2030 projections of GPU pod and rack power consumption (Appendix~\ref{sec:appendix-gpu-projections}) exceed the row power limit for our designs. If a pod or rack exceeds the busbar power limit, cross-row cables can be used to draw power capacity from adjacent rows.

\textbf{Harvesting and decommissioning.}
Racks are often over-provisioned at deployment and later power oversubscribed or \emph{harvested} to better match observed utilization~\cite{kumbhare2021prediction}.
Decommissioned racks release their allocated resources. Lifetimes  are modeled as $\mathcal{N}(\mu_r, \sigma_r)$ using hardware-class-specific parameters (Section~\ref{sec:scheduling}).

\subsection{Placement and Scheduling}\label{sec:placement}
A deployed rack is assigned to a specific row and tile subject to hierarchical resource constraints. If no feasible placement exists in any active hall, a new hall is constructed instantly.
This simplification of the hall commissioning process isolates the effects of the design on the end-to-end metrics, which are subject to noise from demand prediction errors.

\textbf{Placement constraints.}
The power hierarchy is modeled as a tree from substation to row. A placement is feasible if adding a rack does not exceed effective capacity at any ancestor node, where effective capacity is the residual capacity available after enforcing redundancy constraints.

Under distributed $xN/y$ redundancy, effective HA capacity at node $\ell$ is $(y/x)\cdot C_\ell$, while low-availability (LA) racks may use the full capacity $C_\ell$. Cooling, space, and networking resources impose additional row-level constraints. Appendix~\ref{app:placement-formalization} gives the full ancestor-path feasibility condition and demand vector notation.
\emph{Deployment quanta} is modeled as the minimum number of same-SKU racks that must be placed together in one row, consistent with contemporary rack placement mechanisms~\cite{baxi2025onlinerackplacementlargescale}

\begin{figure}
    \centering
    \includegraphics[width=0.95\linewidth]{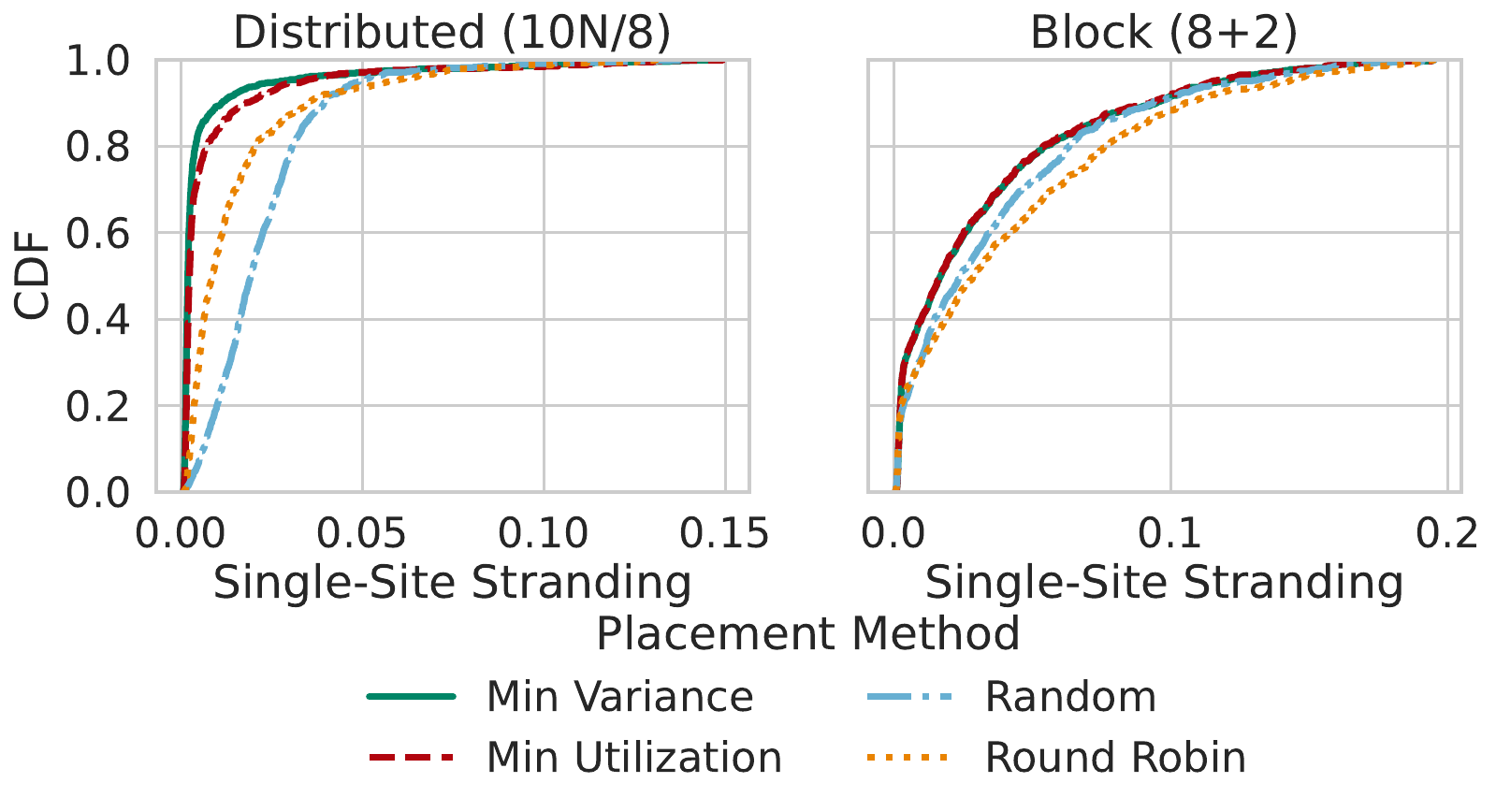}
    \caption{Line-up-level stranding in Monte Carlo simulation of a $10N/8$  and $8+2$ hall populated with storage, compute, and GPU racks under four online placement policies. Variance minimization yields the lowest stranding.}
    \label{fig:placement-policy}
\end{figure}\begin{figure}
    \centering
    \includegraphics[width=0.85\columnwidth]{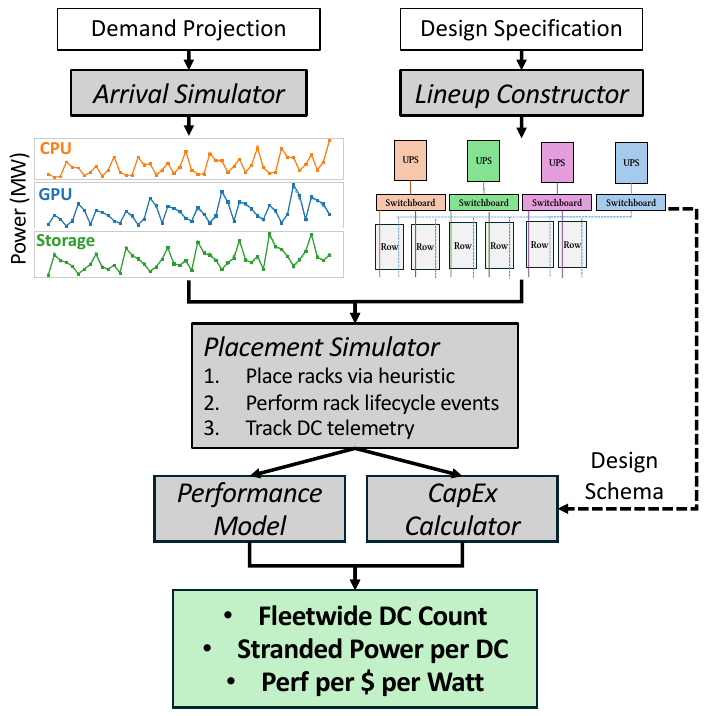}
    \caption{Evaluation pipeline for comparing datacenter power-delivery designs over the deployment lifecycle. Demand projections and design specifications drive a placement simulator, cost model, and performance model, which together produce fleet-level deployability, stranding, cost, and throughput metrics.}
    \label{fig:full-pipeline}
\end{figure}

\textbf{Placement policies.}
Rack placement decisions can have a large effect on stranding.
We evaluate four online heuristics: \textbf{min waste} which uses a best-fit heuristic based on existing placements, \textbf{random}, \textbf{round robin} across rows, and \textbf{variance minimization}. Variance minimization places each deployment with the goal of minimizing imbalance across UPS domains to reduce stranded capacity, especially in distributed-redundant power delivery systems. We observe the lowest stranding from variance minimization; results from a Monte Carlo placement policy comparison is shown in Figure~\ref{fig:placement-policy}; variance minimization is used as the default policy throughout the rest of this paper.

\subsection{Stranding Metrics}
\label{sec:metrics}

A hall may retain unused power, cooling, or space and still be unable to admit another rack because some other constraint binds first. \emph{Stranded capacity} is provisioned capacity that remains unused because it cannot be converted into deployed load under the current placement constraints.

For each resource dimension $m \in \{$power, air cooling, liquid cooling, space$\}$, let
\[
U_t^{(m)} = C_{\mathrm{prov}}^{(m)} - L_t^{(m)}
\]
denote unused provisioned capacity at time $t$, where $C_{\mathrm{prov}}^{(m)}$ is total provisioned capacity and $L_t^{(m)}$ is deployed load. Our stranding metrics isolate the portion of $U_t^{(m)}$ that cannot be used because another constraint binds first. Stranding is therefore multi-dimensional: a row may retain 500\,kW of electrical headroom, but if liquid cooling capacity is exhausted, that power is unavailable to accelerator deployments.

We report two cost metrics. \textbf{Initial \$/MW} is the CapEx of one hall normalized by its nameplate IT capacity (Section~\ref{sec:cost-estimation}). \textbf{Effective \$/MW} measures the infrastructure built per MW of deployed IT:
\[
\text{Effective \$/MW} = \textstyle\left(\sum_{i=1}^{n} K_i\right) / \left(\sum_{i=1}^{n} \hat{P}_i\right)
\]
where $K_i$ is the CapEx of hall $i$ and $\hat{P}_i$ is the final deployed IT MW hosted in hall $i$ at the end of the simulation horizon. The gap between these metrics captures how much provisioned infrastructure fails to translate into deployable load.

\subsection{Rack Placement Simulation}\label{sec:placement-sim}
The framework uses two placement simulators: single-hall for mechanism isolation and fleet-scale for multi-year deployment dynamics.

\textbf{Single data hall.}
Single-hall simulation isolates architectural effects without fleet-level confounders. In each Monte Carlo trial, we instantiate one hall, place racks until 100 consecutive arrivals fail, apply harvesting, and resume placement until another 100 consecutive failures occur. Repeating this process over independently sampled arrival traces yields distributions of stranding and bottleneck behavior. This mode is useful for identifying capacity harmonics and resource-ratio mismatches and iterating on the design prior to running longer fleet simulations.

\begin{figure}
    \centering
    \includegraphics[width=0.85\linewidth]{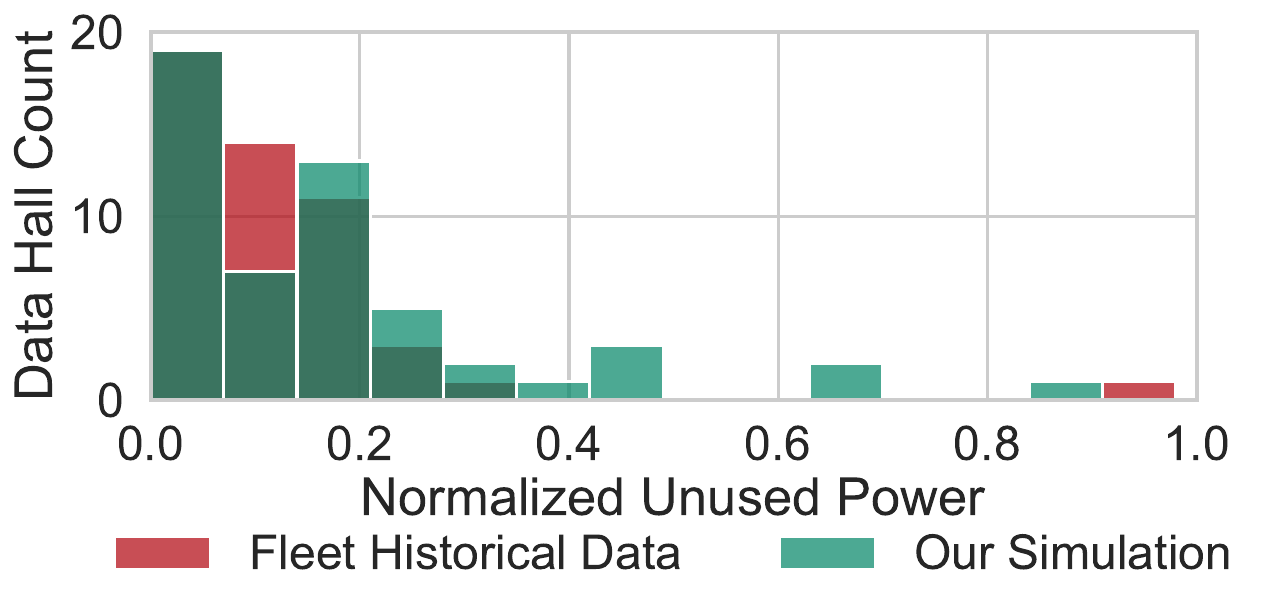}
    \caption{Validation of our simulator against historical rack placements in \azure over 6 years to a subset of both new and mature data halls, and comparing the simulated unused-power distribution to the observed one. Unused power is normalized by the maximum observed value. We report unused rather than stranded power because some halls are not yet saturated.}
    \label{fig:validation}
\end{figure}

\textbf{Datacenter fleet.}
Fleet simulation places racks across multiple halls over a multi-year horizon. This exposes interactions absent in single-hall analysis, including shared arrival budgets across active facilities, cross-hall imbalance, and compounding stranding under heterogeneous arrival sequences. The pipeline has three stages, shown in Figure~\ref{fig:full-pipeline}.

First, an \emph{arrival simulator} (\S\ref{sec:scheduling}) generates monthly rack sequences from long-range demand projections. Second, a \emph{datacenter constructor} instantiates halls from a power-delivery specification. Third, a \emph{placement engine} assigns racks to rows across active halls, opens new halls when existing capacity saturates, harvests power from racks after some time, and decommissions racks at end-of-life. Cost (\S\ref{sec:cost-estimation}) and performance (\S\ref{sec:throughput-model}) models aggregate infrastructure CapEx and workload throughput over the simulation horizon.

The fleet simulation captures interactions between hardware refresh, hall aging, and power-delivery design that do not appear in single-hall analysis. In particular, step changes in GPU rack power can interact with fixed line-up sizes and amplify stranding differences that appear small at single-hall scale.

\textbf{Validation with ground truth.}
We validate the simulator against observed power utilization distributions across data halls over 18 \azure datacenters. 
Historical deployment traces for each datacenter are used to populate simulated halls that match real designs. 
Figure~\ref{fig:validation} compares simulated and observed unused power for racks deployed from 2020 to 2026. We report unused rather than stranded power because some halls are not yet saturated. 
In some cases, $\pm$1 halls are constructed in simulation than in reality, attributed to variance in stranding and harvesting at these sites.
The simulator has a close distributional match to historical observations, with median unused power within 6\% of observed.

\section{Model Instantiation and Experimental Inputs}
\label{sec:implementation}

This section describes the configuration of the evaluation framework with grounded assumptions for demand, hardware evolution, infrastructure cost, and workload throughput. These inputs define the arrival traces, rack and pod characteristics, and comparative performance and cost models used to evaluate how candidate power-delivery designs behave under lifecycle deployment.

\subsection{Arrival Envelopes and Deployment Trace Generation}
\label{sec:rack-demand}

We model lifecycle demand as a time-ordered sequence of rack deployments. These sequences are generated from \emph{arrival envelopes}: monthly capacity targets for each hardware class. The envelopes capture aggregate effects of demand growth, procurement limits, and fleet planning assumptions without committing to a specific forecast.

We consider three hardware classes from 2025 to 2035: accelerators, general compute, and storage. Each class is specified by an initial deployment level, a growth trajectory, and a capacity cap. These determine annual power targets, which are distributed into monthly budgets using seasonality weights stylized after historical \azure procurement cycles, following prior rack-procurement studies~\cite{baxi2025onlinerackplacementlargescale, ishaiefficient}.

\begin{figure}
    \centering
    \includegraphics[width=\columnwidth]{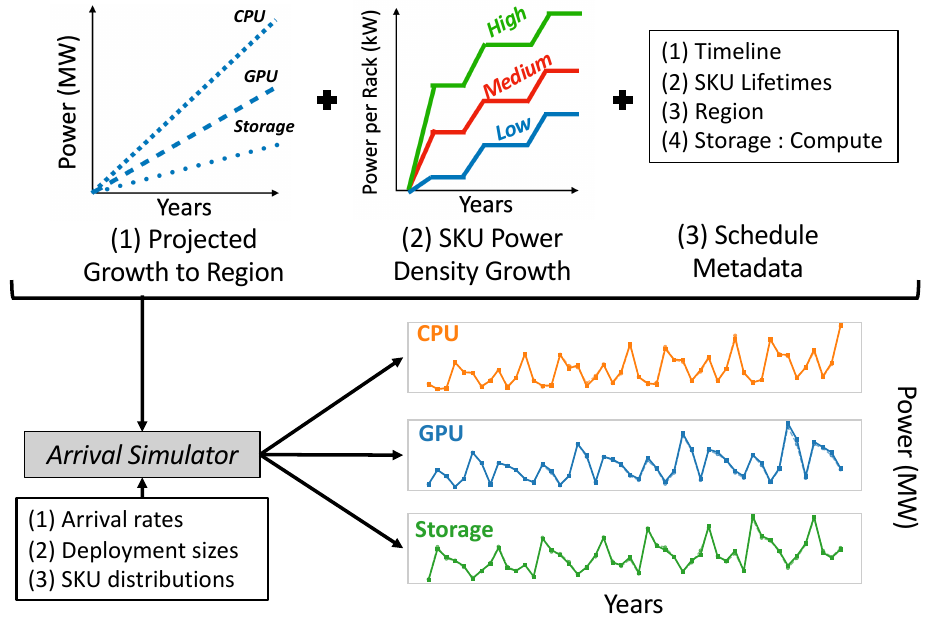}
    \caption{Deployment-trace generation pipeline. Stage~(1) specifies class-level arrival envelopes, stage~(2) assigns per-SKU rack power, and stage~(3) adds lifecycle metadata such as availability tier and retirement time.}
    \label{fig:arrival-sim-pipeline}
\end{figure}

Within each monthly budget, discrete rack arrivals are sampled. Each arrival carries a demand vector $d_r(\tau)$ and relevant lifecycle metadata.
The resulting traces preserve long-range growth trends while capturing the variance and heterogeneity in rack arrivals.

\subsection{Rack Resource Projections and Lifecycle Parameters}
\label{sec:scheduling}
Arrival envelopes determine how much capacity enters the fleet, yet it is necessary to specify how each arriving deployment is assigned rack power, cooling demand, harvesting behavior, and lifetime.

\begin{figure}
    \subfigure[General Compute]{
    \label{fig:compute-hist}{\includegraphics[width=0.48\columnwidth]{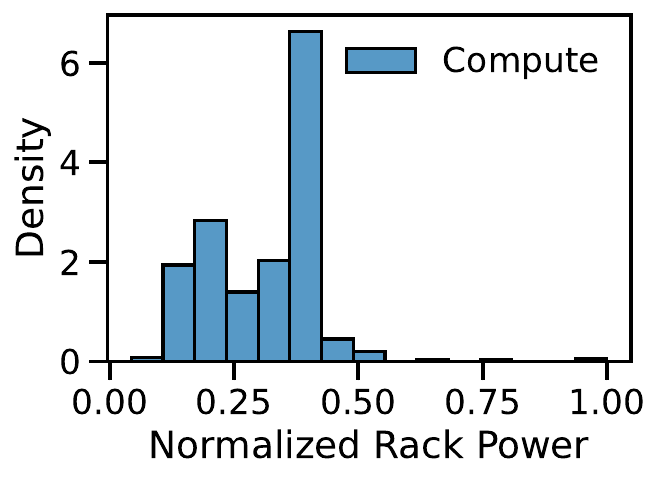}}}
    \subfigure[Storage]{
    \label{fig:storage-hist}{\includegraphics[width=0.48\columnwidth]{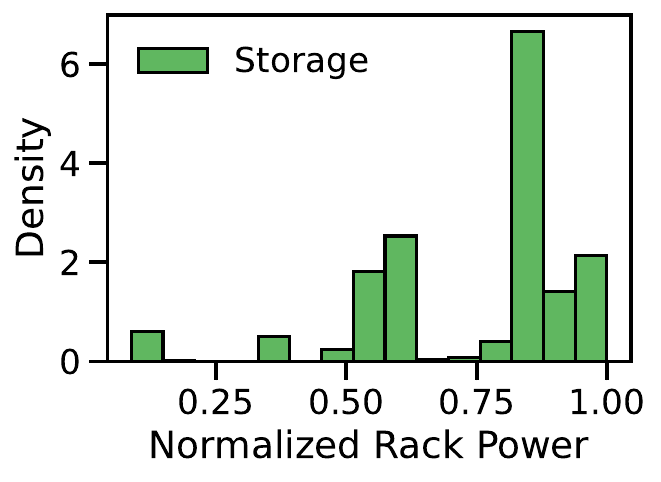}}}
    \caption{Normalized rack-power distributions for \azure general-compute and storage deployments since 2023. These results are clustered into empirical distributions of representative SKU groups for future trace generation.}
    \label{fig:compute-storage-hists}
\end{figure}

\noindent\textbf{General compute and storage SKUs.}
General compute and storage racks exhibit substantial intra-class power variation, as shown in Figure~\ref{fig:compute-storage-hists}. To preserve that heterogeneity, historical \azure rack-power distributions are clustered into representative SKU groups. Each cluster $j$ is defined by a scaling factor $\alpha_j \in (0,1]$ and deployment probability $p_j$.

Given a projected maximum rack power $P_{\max}(\tau, s)$ for hardware class $s$ at time $\tau$, we generate SKU powers as
\begin{equation}
    P_{\textsc{sku},j}(\tau, s) = \alpha_j P_{\max}(\tau, s),
\end{equation}
and sample arriving racks according to $p_j$. This preserves both class-level power growth and the within-class variation observed in past deployments. Appendix~\ref{sec:appendix-cost} gives the non-accelerator power projections.

\noindent\textbf{Accelerator rack case studies}
Accelerator rack power is more tightly coupled to hardware generation and system form factor, so it is modeled explicitly by year and scenario.
\footnote{We capture broader industry trends: rising accelerator package power, denser rack integration, larger local high-bandwidth accelerator domains, and increasing sensitivity to inter-domain communication.} 
For each year and growth scenario, a rack TDP anchored to public hardware disclosures is assigned and extended using the per-package TDP growth model in Appendix~\ref{sec:appendix-projections}.

Our near-term case studies use dense rack-scale systems. For later years, we include a denser rack case to capture the possibility that future accelerator deployments arrive at higher effective rack power~\cite{huang-kyber-600kw,nvidia-system-scaling}. We also distinguish \emph{rack-scale} accelerator arrivals from \emph{pod-scale} arrivals. Rack-scale arrivals are single rack-local deployment units. Pod-scale arrivals represent multiple accelerator racks deployed together on a shared low-latency pod fabric, with aggregate demand equal to the sum across constituent racks. Across all hardware classes, these projections draw on public vendor disclosures and industry reports~\cite{nvidia-rubin, nvidia-dsx-blueprint, nvidia-rubin-ultra, intel-xeon-tdp, delloro-ai-compute, datacenter-processors-2025, nvidia-system-scaling, semianalysisrubin}.

\noindent\textbf{Harvesting and lifetimes}
Harvesting and retirement determine when occupied capacity returns to the fleet. One year after deployment, one can optionally harvest up to 15\% of provisioned power and cooling from storage and general-compute racks and up to 10\% from accelerator racks, approximating median historical rates observed at \azure. Rack lifetimes are drawn from $\mathcal{N}(7,1)$ years for storage and general compute and $\mathcal{N}(5,0.5)$ years for accelerator rack and pod deployments.

\subsection{Component-Based Infrastructure Cost Model}
\label{sec:cost-estimation}
The simulator determines how many halls of a given topology must be built to serve a deployment stream. To compare those halls on CapEx rather than nameplate capacity alone, we use a topology-based, per-component cost model. This preserves hall-specific effects while evaluating all designs under the same cost assumptions.

Datacenter construction costs are proprietary, vendor- and site-dependent, hence our goal is not to predict exact project cost, but rather develop a comparative model. Published Tier~III/IV surveys place non-IT infrastructure cost at roughly \$7--12\,M/MW~\cite{CushWakeCostGuide_2025, AccutechBuildCost_2024, barroso2019datacenter, turnerandtownsend2024index, dgtlinfra2024costs, thundersaid2024economics}. \$10\,M/MW is a reference baseline, consistent with Schneider Electric's per-rack scaling model~\cite{mccarthyavelar2016ups, uptimeinstitute2023costs} and recent industry surveys~\cite{CushWakeCostGuide_2025, dgtlinfra2024costs}. This baseline includes electrical, mechanical, building-shell, and soft construction costs, includes a high liquid-cooling share for accelerators~\cite{schneider2020wp282}, and excludes IT equipment.

\noindent\textbf{Per-component costing}
Where published equipment pricing is available, we use documented ranges. The remaining electrical budget is allocated across generators, transformers, switchgear, and transfer switches. Table~\ref{tab:component_costs} summarizes all per-component costs. Because different power-delivery topologies imply different counts and ratings of UPS modules, switchboards, busway runs, and transfer equipment, the same nameplate IT capacity can map to different hall CapEx. These hall-level costs feed directly into the initial and effective \$/MW metrics defined in Section~\ref{sec:metrics}.

\subsection{Workload Throughput Model}
\label{sec:throughput-model}

Infrastructure differences matter only insofar as deployed hardware is performant. Therefore the infrastructure simulation is complemented with a workload model that maps deployed accelerator capacity to comparative inference throughput. We focus on LLM inference, which we expect to account for a substantial share of utilization in datacenters not specialized for training. The model is comparative rather than fully operational: it estimates how accelerator generation, rack density, local interconnect scale, and inter-domain communication translate into delivered tokens per second under representative inference workloads.

Inference throughput is modeled as a mixture-of-experts (MoE) architecture\footnote{Many current frontier models use MoE architectures.}~\cite{deepspeedmoe2022, tithi2025scalingintelligencedesigningdata} running on accelerator deployment $D$. Hardware inputs such as local accelerator-domain size, HBM capacity, HBM bandwidth, and compute throughput are drawn from public hardware disclosures and from extrapolations used in prior TCO analyses~\cite{stojkovic2025rearchitectingdatacenterlifecycleai}. The two standard inference phases are considered separately: \emph{prefill}, which processes the input prompt in parallel and is typically compute-bound~\cite{spad, prefillonly}, and \emph{decode}, which generates tokens autoregressively and is often limited by HBM bandwidth and KV-cache movement~\cite{vllm, dynamollm, spad}.

\noindent\textbf{Phase bottlenecks.}
For model $m$, deployment $D$, and phase $\phi \in \{\text{pre}, \text{dec}\}$, throughput is limited by the slowest of three resources: compute, HBM bandwidth, and communication:
\begin{equation}
    \mathrm{TPS}^{\phi}(m,D)
    =
    \min\!\left(
        \frac{F_D}{\mathcal{C}^{\phi}(m)},
        \frac{B_D^{\mathrm{HBM}}}{\mathcal{M}^{\phi}(m)},
        \frac{1}{T_{\mathrm{comm}}^{\phi}(m,D)}
    \right).
    \label{eq:phase-bottleneck}
\end{equation}
Here $\mathcal{C}^{\phi}(m)$ and $\mathcal{M}^{\phi}(m)$ are the per-token compute and memory costs implied by model $m$, and $T_{\mathrm{comm}}^{\phi}(m,D)$ captures tensor-parallel (TP) and expert-parallel (EP) communication on deployment $D$. A fixed serving batch size of $B=256$ is used to represent a high-throughput operating point. Prefill and decode differ in both memory traffic and communication structure: prefill amortizes weight traffic across the prompt batch, whereas decode depends on active weights and growing KV-cache reads over the generated sequence. Appendix~\ref{sec:appendix-perf-model} gives the full formulation.

\noindent\textbf{Communication model.} Communication depends on how much EP traffic remains within a local high-bandwidth accelerator domain and how much spills across domains onto the cluster fabric. Larger local domains keep more traffic on the fast on-package or rack-local interconnect. Smaller domains, or larger models, incur more inter-domain communication. Appendix~\ref{sec:appendix-capacity} and Appendix~\ref{sec:appendix-perf-model} provide the corresponding capacity and model details.

The throughput model provides the performance side of the comparison in Figure~\ref{fig:space-complexity}. It lets us ask how much local accelerator scale improves inference throughput per watt, while the placement simulator determines whether the resulting rack or pod can actually be deployed.

\section{Evaluation of Datacenter Designs}
\label{sec:eval}

\begin{figure} \subfigure[GPU rack-scale systems]{ \label{fig:rack-density}{\includegraphics[width=0.4\columnwidth]{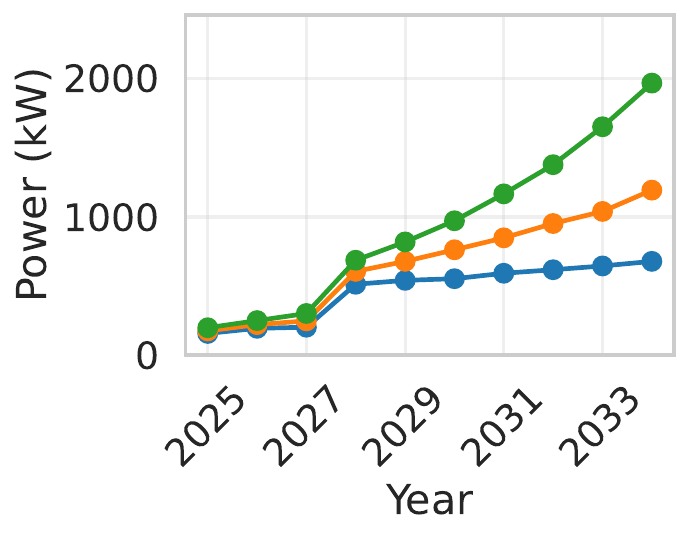}}} \subfigure[GPU pod-scale deployments]{ \label{fig:pod-density}{\includegraphics[width=0.57\columnwidth]{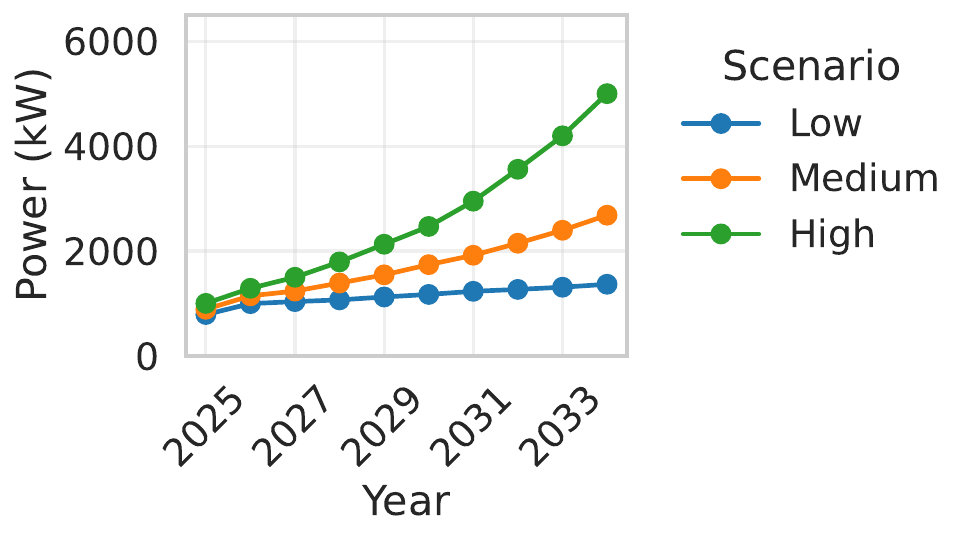}}} \subfigure[Compute]{ \label{fig:compute-density}{\includegraphics[width=0.37\columnwidth]{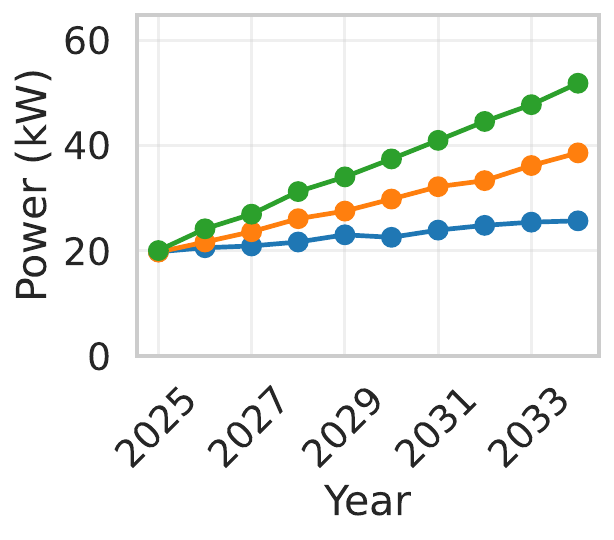}}} \subfigure[Storage]{ \label{fig:storage-density}{\includegraphics[width=0.56\columnwidth]{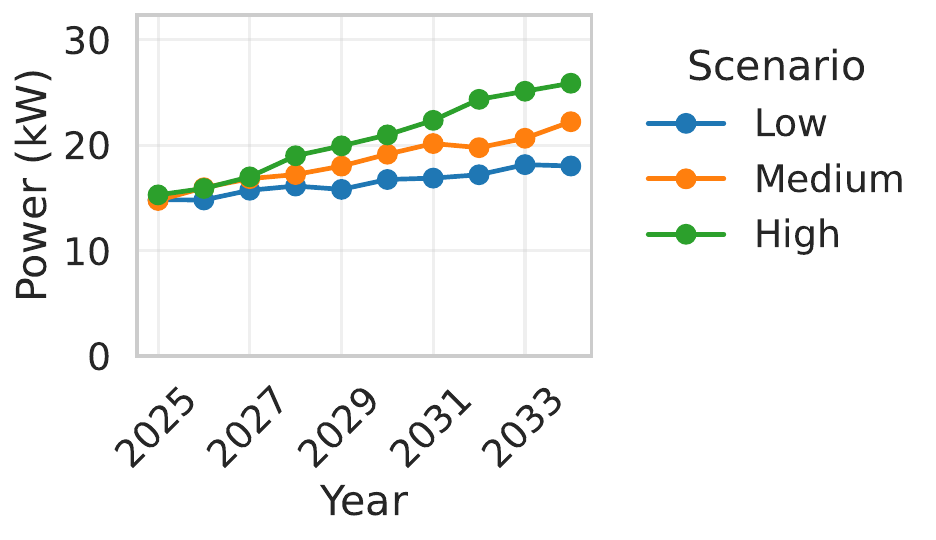}}} 
\caption{Projected power-density trajectories for GPU racks, GPU pods, CPU compute racks, and storage racks.} 
\label{fig:power-schedules} \end{figure}

We evaluate how commissioning metrics evolve for different designs as hardware TDP grows and halls partially fill with deployments. The evaluation seeks to answer four questions:

\begin{enumerate}
    \item How do designs with similar installed HA capacity and base \$/W compare over an 8-year horizon in terms of cost, power capacity, and workload throughput?
    \item Does fleet-scale stranding follow the topology-specific mechanisms from Section~\ref{sec:mechanism-isolation}?
    \item Can operational levers reduce stranding enough to change design rankings?
    \item When do larger GPU pods deliver enough throughput gain to justify their larger placement quanta?
\end{enumerate}

\subsection{Experimental Setup}

\begin{table}[!t]
\centering
\caption{Evaluation setup. We compare redundancy 
  topologies under three GPU power-density trajectories 
  and pod compositions, holding demand, infrastructure 
  granularity, and operational policy fixed.}
\label{tab:eval-setup}
\resizebox{\columnwidth}{!}{
\begin{tabular}{@{}lp{5.2cm}@{}}
\toprule
\textbf{Parameter} & \textbf{Setting} \\
\midrule
\multicolumn{2}{@{}l}{\textit{Comparison axes 
  (varied)}} \\[2pt]
Redundancy topology 
  & $4N/3$ vs.\ $3{+}1$ (7.5\,MW HA) vs.\newline
    $10N/8$ vs.\ $8{+}2$ (20\,MW HA) \\
GPU TDP trajectory 
  & Low / Medium / High 
    (Fig.~\ref{fig:power-schedules}) \\
GPU deployment unit 
  & Single rack;
    pods of 3--7 racks \\
\midrule
\multicolumn{2}{@{}l}{\textit{Infrastructure 
  (fixed)}} \\[2pt]
Buildout horizon 
  & 2026--2034 \\
Cumulative IT demand 
  & 10\,GW: 6.0 GPU, 2.8 compute,\newline
    1.2 storage \\
Electrical granularity 
  & 2.5\,MW UPS; 625\,kW LD row;\newline
    2.5\,MW HD row \\
HD and LD row count
  & 3:2 ratio of LD to HD rows;\newline
  (Appendix~\ref{sec:appendix-rows}). \\
\bottomrule
\end{tabular}}
\end{table}

Table~\ref{tab:eval-setup} summarizes the evaluation setup. The central question is whether designs with similar installed HA capacity remain similar once evaluated over a multi-year deployment lifecycle. We therefore match  designs by nameplate capacity---$4N/3$ versus $3{+}1$ at  7.5\,MW, and $10N/8$ versus $8{+}2$ at 20\,MW---and subject them to the same 10\,GW demand  stream.\footnote{Not indicative of any vendor, hyperscaler, or specific projection.} GPU deployments follow three power-density trajectories (Fig.~\ref{fig:power-schedules}); compute and storage use the Medium trajectory throughout.

We report three classes of metrics at increasing levels of fidelity. Commissioning metrics (installed HA~MW, base\$/W) capture what a design looks like at construction. Lifecycle metrics (P90 site stranding, effective \$/W, halls built) capture what it delivers after years of arrivals, harvesting, and retirements. Workload metrics (inference TPS/W) capture whether deployed hardware translates into useful throughput. For the workload study, we evaluate MoE inference spanning 0.6\,T to 401\,T parameters (as shown in Table~\ref{tab:model-suite}), from models that fit within a single rack-local accelerator domain to models where expert-parallel communication benefits from pod-local placement.

We also sweep GPU share from 40\% to 80\% of total fleet power. The qualitative design ranking is stable across this range; higher GPU shares widen the gap because fewer low-density racks remain to absorb residual capacity fragments.

\subsection{How Static Metrics Change with Rack TDP Growth}
\label{sec:eval-static-fails}

We first ask whether static commissioning metrics remain predictive as accelerator power grows over the facility lifetime. If they do, then designs with similar nameplate MW and similar base \$/W should remain close when evaluated over multi-year deployment traces.

\begin{figure}
    \centering
    \includegraphics[width=\columnwidth]{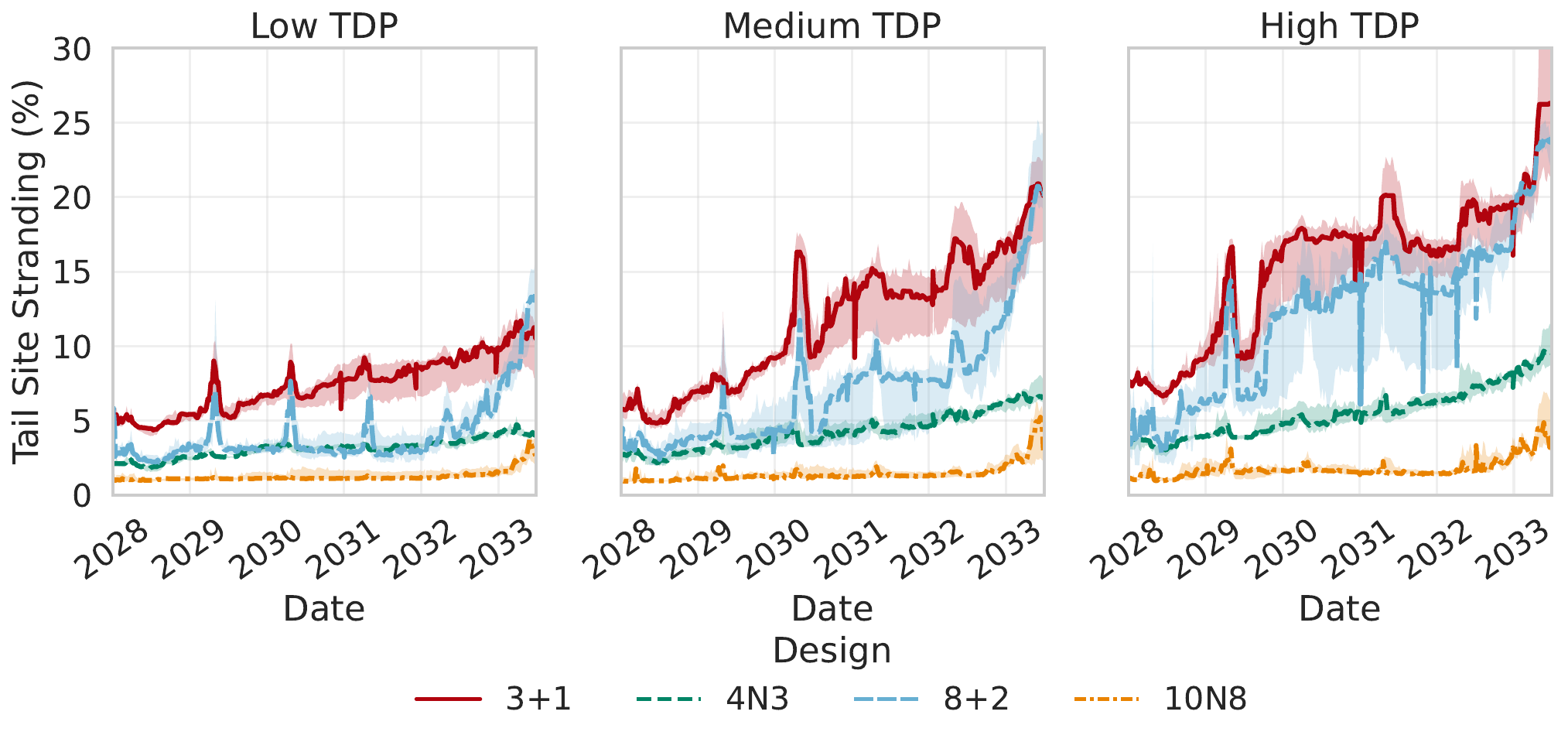}
    \caption{Tail (P90) site stranding over time for block-redundant (3+1, 8+2) and distributed (4$N$/3, 10$N$/8) designs under Low, Medium, and High GPU TDP projections. Lines show the median across pod compositions (3--7 racks); bands span the min--max range. Designs that appear similar under static capacity metrics separate once evaluated over the deployment lifecycle.}
    \label{fig:time-series-power-sensitivity}
\end{figure}

\noindent\textbf{Lifecycle deployability separates designs.} Figure~\ref{fig:time-series-power-sensitivity} shows tail site stranding over time under low, medium, and high GPU power density trajectories from Figure~\ref{fig:power-schedules}. The key result is not just that stranding rises with TDP, but that designs with similar static HA capacity separate materially once future deployments must be placed into partially filled halls. Under the High trajectory, the 3+1 design exceeds 20\% tail stranding by 2033, while 4$N$/3 remains below 10\%. Increasing the number of line-ups helps both families, but even as 8+2 improves over 3+1, it still strands more capacity than 4$N$/3 under the more aggressive projections. 

The reversal arises because future deployments must fit into the residual headroom left by earlier ones. What matters is not aggregate remaining slack, but whether that headroom remains in the right UPS domains, rows, and line-ups after years of arrivals, harvesting, and retirements. Rising GPU TDP makes this distinction visible because larger deployment quanta consume a larger fraction of each electrical component in the distribution network.

\begin{figure}
    \centering
    \includegraphics[width=\linewidth]{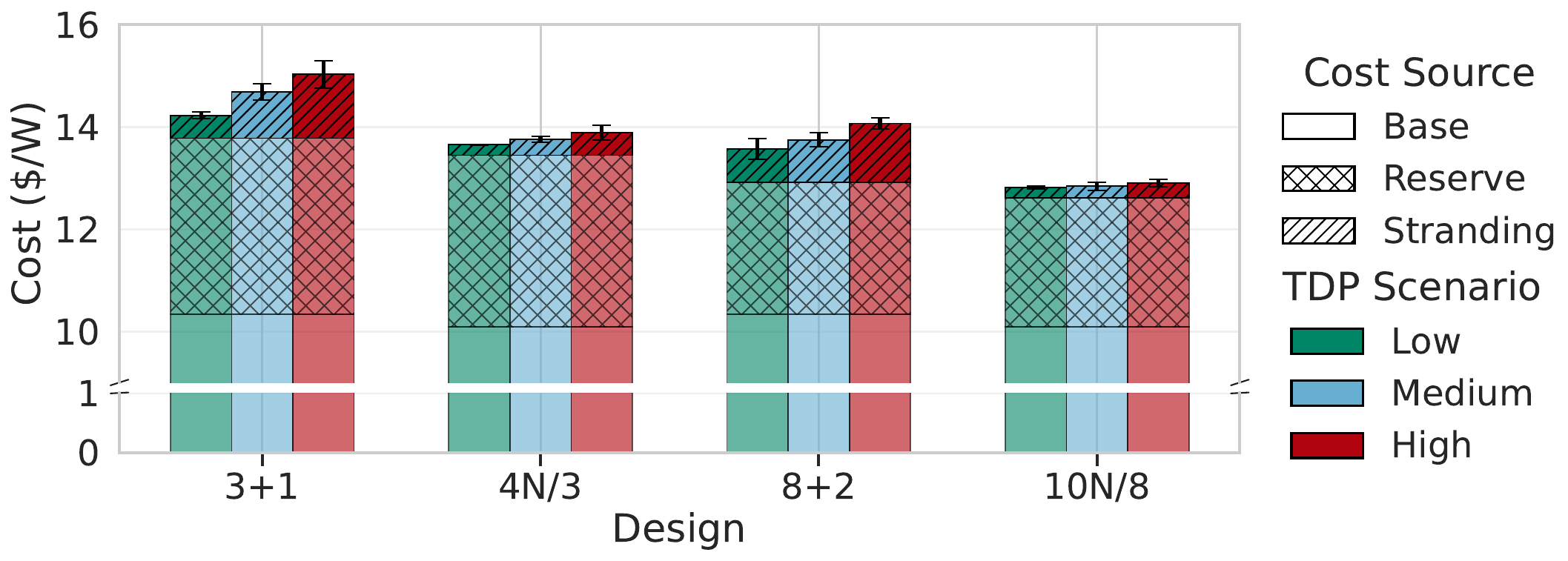}
    \caption{Incremental effective cost above each design's base \$/W. Bars decompose this excess into reserve cost and stranding-induced cost. Error bars show standard deviation across pod compositions. The main moving term is the cost of stranded capacity, not the nominal cost of reserve.}
    \label{fig:stranded-tax-fixed}
\end{figure}

\noindent\textbf{The deployability gap changes cost metrics.}
This separation is not only a utilization effect. As stranding rises, additional halls must be built to serve the same IT load, which appears directly as higher effective cost. Figure~\ref{fig:stranded-tax-fixed} decomposes effective cost above each design's base \$/W into reserve cost and stranding-induced cost. All designs begin with similar base costs, and reserve varies only modestly with redundancy architecture. The main source of variation is instead stranded capacity: infrastructure that was provisioned but cannot be converted into deployed IT load.

As GPU power rises, designs with worse lifecycle deployability convert more provisioned capacity into unusable fragments, which raises effective cost above the base. This effect is largest in 3+1, smaller in 4$N$/3, and remains muted in 10$N$/8 across all three TDP projections. The implication is that designs that begin with similar static costs do not remain similar once evaluated by the deployable capacity they preserve over the fleet lifecycle.

This is why designs separate along the cost axis in Figure~\ref{fig:space-complexity}. The difference is not only what the hall costs to build, but how much of that built capacity remains deployable after years of arrivals, oversubscription, and retirements.

\subsection{Topology Mechanisms of Lifecycle Stranding}
\label{sec:eval-mechanism-validation}

We have shown that static installed-capacity metrics mis-rank designs as hardware evolves. This subsection asks whether the resulting lifecycle stranding follows the topology-specific mechanisms identified in Section~\ref{sec:mechanism-isolation}.

\begin{figure}
    \centering
    \includegraphics[width=\linewidth]{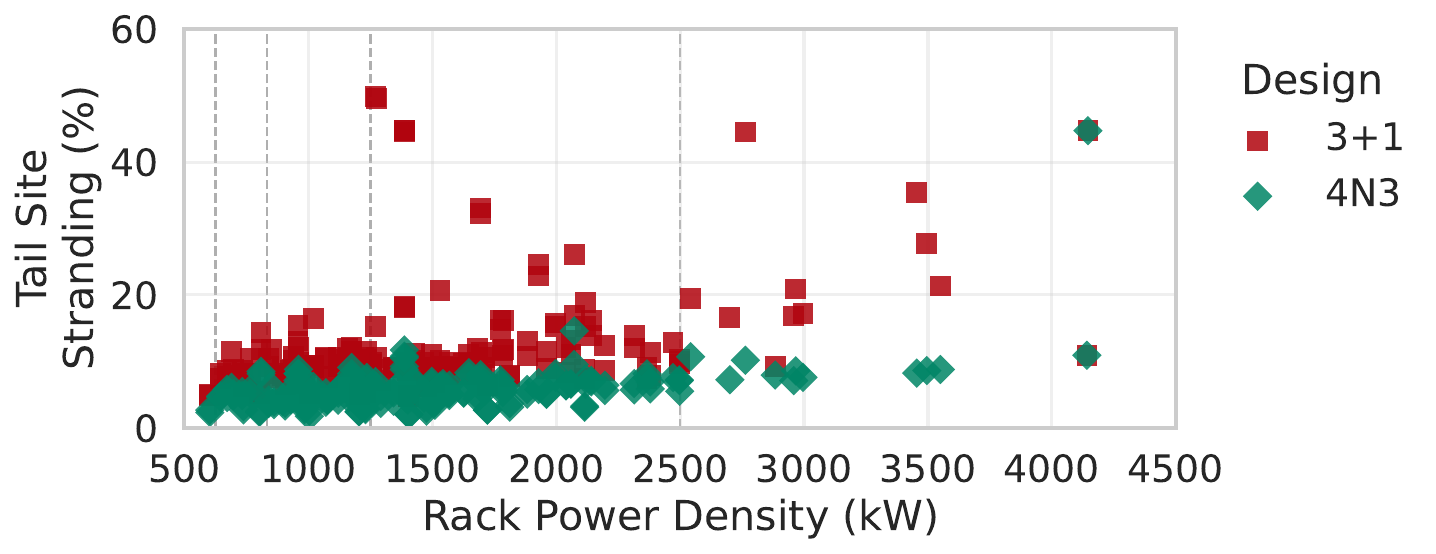}
    \caption{P90 tail stranding versus effective per-domain deployment power for 3+1 and 4$N$/3 across all GPU TDP scenarios and pod compositions. Dashed vertical lines mark 2.5\,MW UPS-block quantization thresholds, around which 3+1 exhibits pronounced stranding increases.}
    \label{fig:effective-tdp-sawtooth}
\end{figure}
Figure~\ref{fig:effective-tdp-sawtooth} plots P90 tail stranding against effective per-domain deployment power across GPU TDP scenarios and pod compositions. The pattern matches the structural distinction from Section~\ref{sec:mechanism-isolation}. In 3+1, high-stranding points cluster near the dashed $C/q$ thresholds of a 2.5\,MW UPS block. Just above such a threshold, one fewer deployment fits, and the residual capacity becomes provisioned but undeployable, consistent with the quantization effect in Figure~\ref{fig:harmonic-stranding}. The 4$N$/3 design degrades differently. Stranding still rises with deployment power, but not around discrete thresholds. Instead, it increases as larger deployments make it harder to satisfy the multiple-parent redundancy constraint.

Figure~\ref{fig:effective-tdp-sawtooth} also demonstrates that at lower TDPs,  block redundancy is an effective design. At commissioning and at lower rack densities, block and distributed designs can look comparable, and block designs retain practical advantages for their simple failure modes. The gap appears when high-power AI deployments make placement granularity a first-order constraint. 

This also explains the variation across pod compositions. Changing pod size changes the effective deployment quantum seen by the hierarchy. Some quanta align poorly with upstream electrical ratings and produce abrupt jumps in undeployable capacity. Others mainly tighten placement feasibility and produce a more continuous loss of deployability. The fleet-scale behavior therefore follows the same topology-dependent mechanisms seen in our analysis.

\subsection{Operational Levers for Reducing Stranding}
\label{sec:eval-levers}

A natural question is whether operational levers that apply comparably across these hierarchies can recover enough lifecycle loss to change design performance. If so, adjusting deployment quanta or harvesting should materially reduce cost and narrow the gap across designs.

\begin{figure}
    \centering
    \includegraphics[width=\columnwidth]{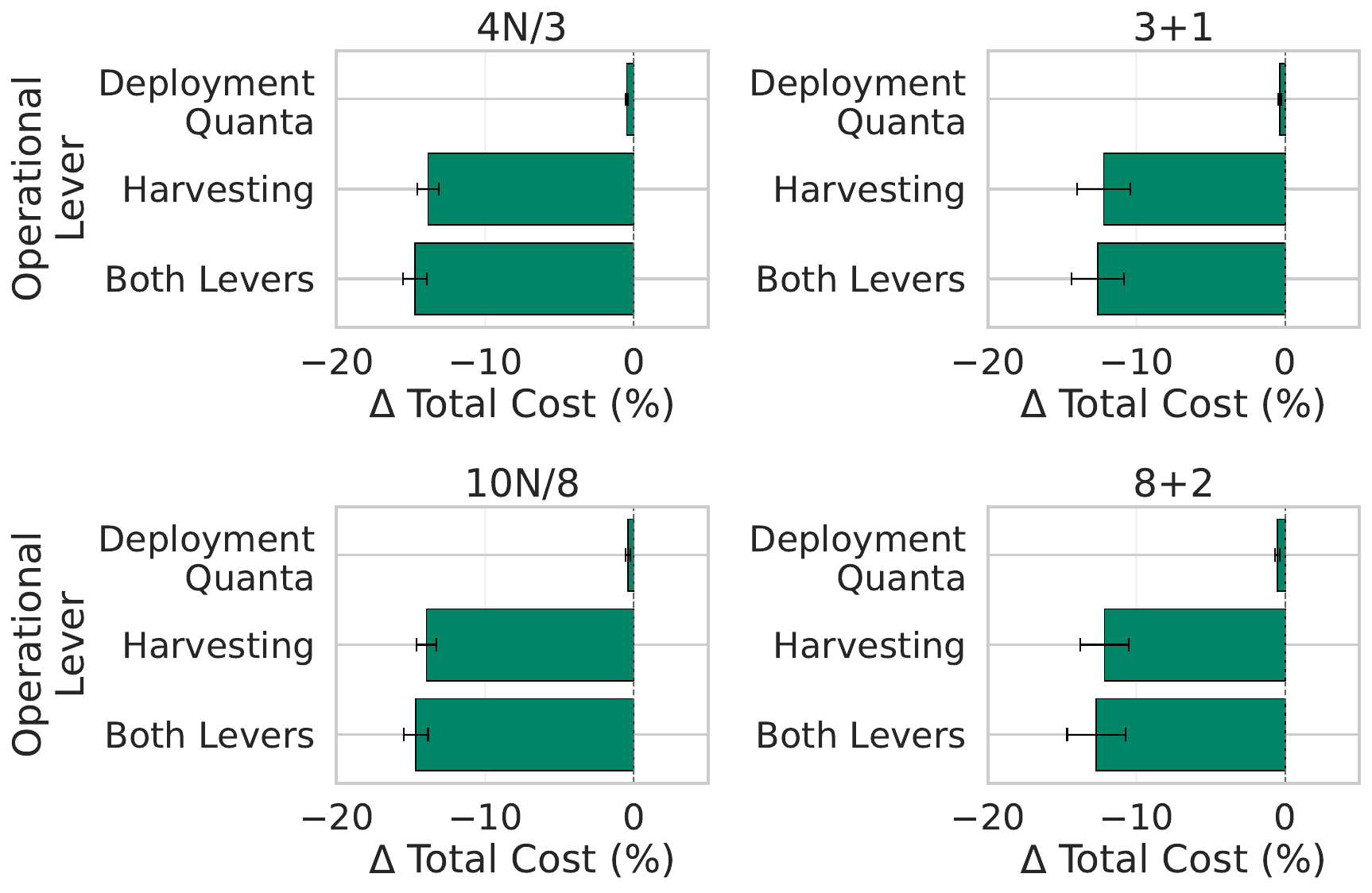}
    \caption{Change in total cost relative to the baseline fleet under the best setting from each operational lever family. Negative values indicate cost savings. Tuning reduces some costs, but does not change design outcomes.}
    \label{fig:operational-levers}
\end{figure}

Figure~\ref{fig:operational-levers} reports the largest cost reduction achieved by each lever family relative to the baseline of 10 racks per deployment quantum and no harvesting. Smaller deployment quanta reduce some admission failures because they make low-density racks easier to pack. This effect is more pronounced in single-site packing, but at fleet scale it translates into only about 1\% fewer data halls because large accelerator racks remain difficult to place.

Harvesting returns occupied power and cooling to the fleet, opening some additional placement opportunities, but the gains are modest and do not change the design ranking. More generally, operational tuning can reduce imbalance, return some capacity, or make individual arrivals easier to place, but it does not change the underlying placement structure imposed by hierarchical high-availability constraints. Service-model relaxations such as low-availability tiering~\cite{zhang2021flex} can reclaim reserve capacity, but they alter the comparison: in distributed designs, reserve is shared across active parents, whereas in block designs, low-availability load would sit on primary line-ups without locally provisioned failover. We therefore treat such relaxations as orthogonal to the structural question studied here. 

Even with these levers, once the hierarchy’s constraints bind, the remaining slack is structurally hard to use. Policy can improve utilization of residual capacity, but it does not materially change which future racks and pods the hierarchy can admit.

\subsection{When do GPU pod networking gains survive deployability constraints?}
\begin{figure}
\centering
\includegraphics[width=\linewidth]{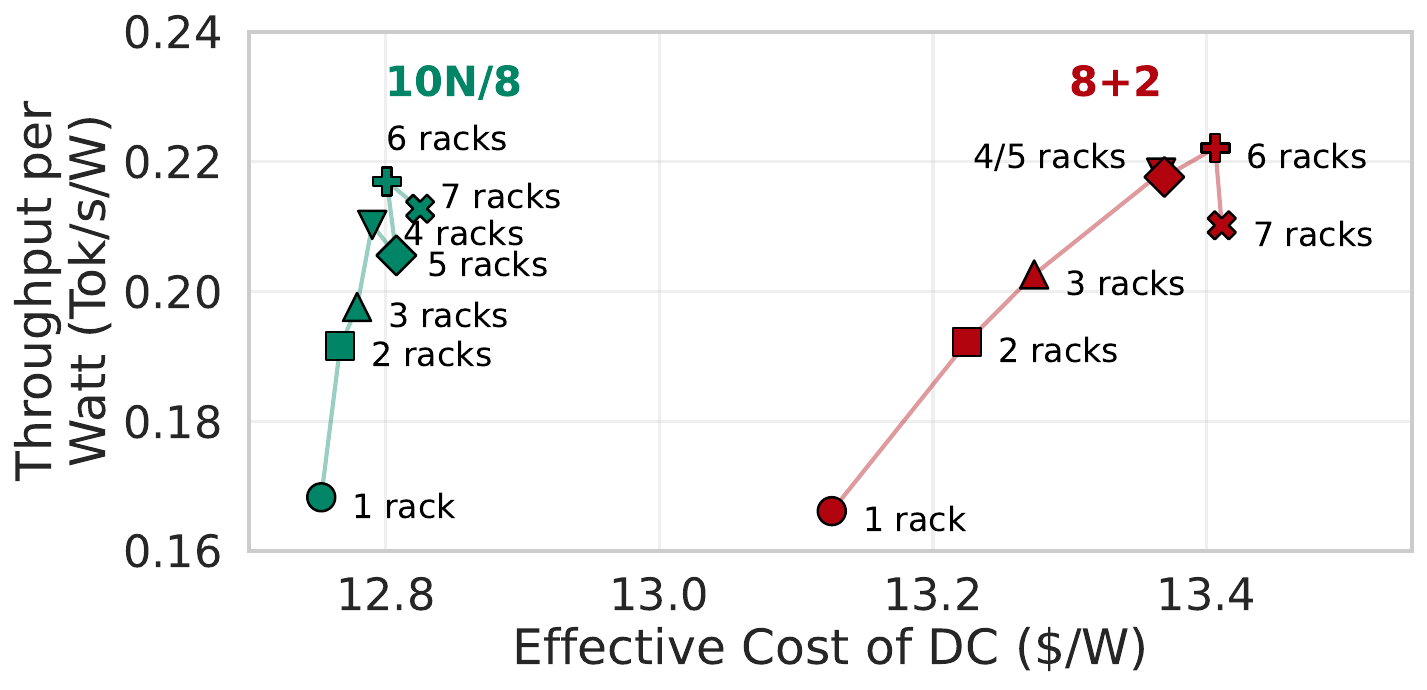}
\caption{Effective fleet cost versus power-normalized throughput under High GPU TDP growth for MoE-132T. Larger pods increase throughput but also raise effective cost by increasing deployment granularity. 10$N$/8 preserves more of the throughput gain at fleet scale because it admits larger quanta with less deployability loss.}
\label{fig:cost-throughput-tradeoff}
\end{figure}

Larger pods improve serving efficiency by keeping more EP communication within a local high-bandwidth domain, but they also arrive as coarser indivisible placement quanta~\cite{nvidia-800vdc,semianalysisrubin}. The question is whether that serving gain survives the lifecycle deployability penalty imposed by the hierarchy.

Figure~\ref{fig:cost-throughput-tradeoff} makes that tradeoff explicit for two designs serving an MoE-132T workload (scaled factors of DeepSeek-R1's model parameters~\cite{deepseek}). Moving to larger pods shifts both designs upward and to the right. The upward movement is the serving-side effect: more traffic remains within the local domain, so throughput per watt increases. The rightward movement is the infrastructure-side effect: larger quanta are harder to place into partially filled halls, so effective fleet cost rises. In 10$N$/8, larger pods preserve more of their serving benefit at fleet scale. In 8+2, the same performance gain is offset more heavily by the cost of admitting a coarser quantum. The difference is not the serving model, but the hierarchy's ability to absorb larger placements without giving up as much deployable capacity.

To expose the crossover directly, define \emph{pod payoff} relative to a single-rack baseline:
\[
\text{Pod Payoff}
=
\frac{1 + \Delta \text{TPS/W}}
     {1 + \Delta \text{Cost}}
- 1,
\]
where $\Delta \text{TPS/W}$ is the fractional throughput-per-watt gain from better networking and $\Delta \text{Cost}$ is the fractional increase in effective fleet cost induced by the larger placement quantum. Positive payoff means the serving gain exceeds the deployability cost. Negative payoff means the communication benefit is outweighed by lifecycle deployment cost.

\begin{figure}
\centering
\includegraphics[width=\linewidth]{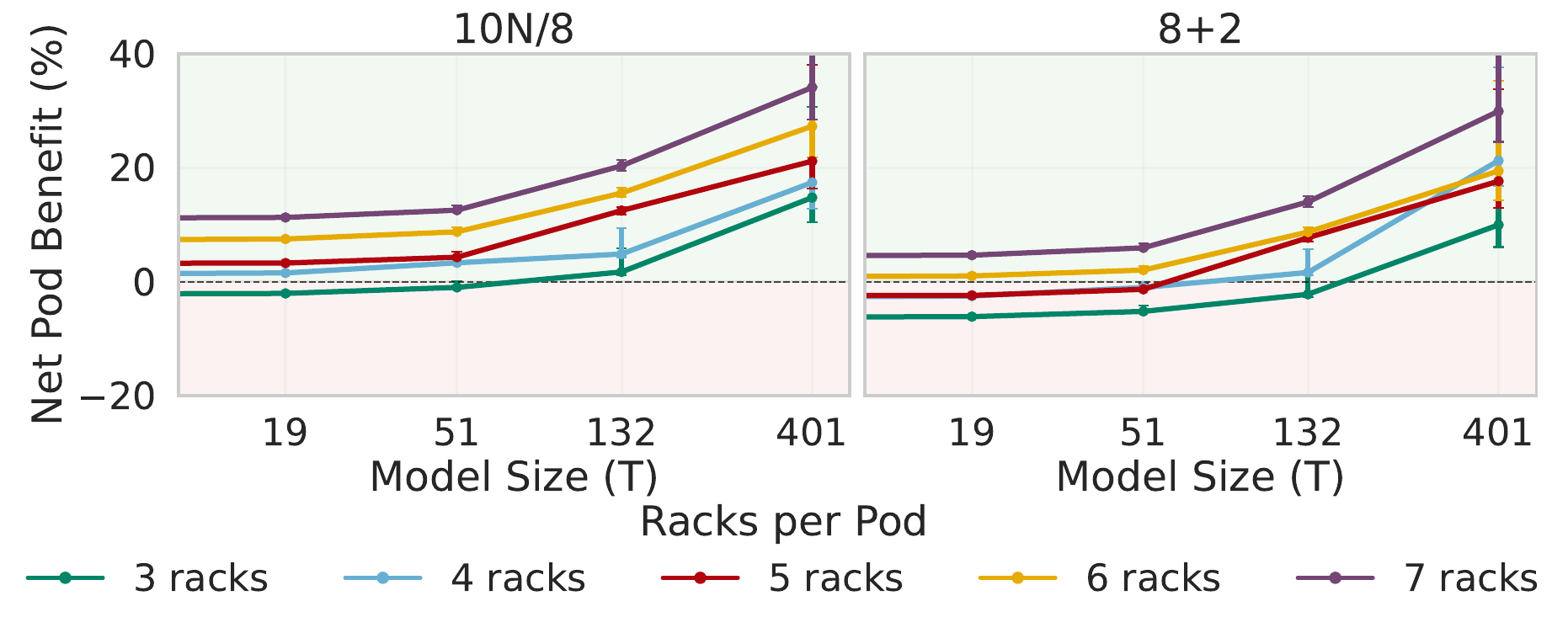}
\caption{Pod payoff across model sizes for 10$N$/8 and 8+2. Larger pods help only once their communication benefit exceeds their deployability cost. The later crossover in 8+2 reflects its larger deployability penalty.}
\label{fig:pod-payoff}
\end{figure}

Figure~\ref{fig:pod-payoff} shows that the crossover depends on both workload and hierarchy. For smaller models, most communication is already contained within a rack-scale domain and pods have little to offer for serving throughput, but still incur a placement penalty, so payoff remains near zero or negative. As model size grows, more EP traffic spills across domains, and payoff becomes positive.

The crossover is also topology-dependent. In 10$N$/8, payoff becomes positive earlier and rises more uniformly with pod size because the hierarchy admits larger quanta with less deployability loss. In 8+2, the same pod sizes cross divisibility and placement thresholds sooner, leaving less net benefit after lifecycle deployment. What changes across designs is how much of the serving gain remains after years of arrivals into a constrained hierarchy.

Pod efficiency therefore helps only when two conditions hold: (1) the workload is communication-limited enough to benefit from a larger local domain, and (2) the hierarchy is fine-grained enough to admit the deployment quantum without stranding that gain. These conditions are intuitive, and our framework quantifies an estimate for cross-over points for the benefit of pods across model size that vary across datacenter designs. Appetite for the high end of future accelerator power densities may not depend solely on theoretical improvement in hardware performance, but also on how efficiently a datacenter can host them.

\section{Related Work}

\textbf{Datacenter design and rack placement.}
Prior work studies redundancy schemes, capacity provisioning, and rack placement, but usually treats the electrical hierarchy as fixed. Industry guidance and systems texts describe distributed ($xN/y$) and block ($N{+}y$) redundancy and compare them mainly on first-order cost, complexity, and maintainability~\cite{fan2007power, hamilton, torell2016n1cost, ASCO_Redundancy_2019, SchneiderWP110_2025, SchneiderWP61_2019}. A separate line of work formulates rack placement and expansion as online or stochastic packing, improving utilization through better placement policies, migration, or lifetime-aware allocation~\cite{baxi2025onlinerackplacementlargescale, ishaiefficient, half-bin, cohen2017overcommitment, mellou2024binpacking, barbalho2023vmlifetime, flux}.  These systems optimize where to place load, not whether the hierarchy can admit it. Our work complements this literature by exposing how, as deployment quanta consume a meaningful fraction of rows, line-ups, or UPS blocks, power delivery design can be the dominant source of inefficiency.

\textbf{Datacenter power management and oversubscription.}
A large body of work improves utilization within an existing hierarchy through power capping, DVFS, storage, and oversubscription~\cite{argo, dynamo2016wu, li2019capmaestro, raghavendra2008struggles, kumbhare2021prediction, kontorinis2012managing, wang2012energystorage}. Other systems reclaim reserve or underused capacity within a fixed hierarchy by relaxing service guarantees or exploiting utilization headroom~\cite{bashir2021overcommitment, piga2024expanding, meta2024dynamic, zhang2021flex}. Our analysis is orthogonal, as we characterize how hierarchical structure shapes deployability under a fixed service model.

\textbf{AI infrastructure and power delivery.}
Recent work shows that AI training and inference change datacenter requirements through higher rack power, different temporal behavior, and tighter coupling between compute, cooling, and interconnect~\cite{tapas, polca, dynamollm, chung2024reducing, bianchini2024datacenter}. This has prompted new discussions of AI-oriented facility design, including denser cooling integration and alternatives such as 800V DC distribution~\cite{ocp-summit-2025, stojkovic2025rearchitectingdatacenterlifecycleai, nvidia-800vdc}. The closest prior work~\cite{stojkovic2025rearchitectingdatacenterlifecycleai} evaluates AI datacenters through aggregate hardware lifecycle cost; our works provides a complementary perspective on AI datacenter infrastructure.

\section{Conclusion}

AI accelerator growth changes power-delivery design from a commissioning problem into a lifecycle deployability problem. A hall can retain substantial provisioned power and still be unable to admit future racks or pods because residual capacity is fragmented across the hierarchy. Our results show that this gap is large enough to change design rankings: topologies with similar installed capacity and base cost diverge as deployment quanta grow, with stranded capacity becoming the dominant source of effective cost. The design objective for AI datacenters is therefore not installed megawatts, but deployable capacity over time. Power hierarchies should be evaluated by how much useful accelerator capacity they continue to admit across hardware generations, placement histories, and workload requirements. As AI facilities scale, this gap becomes a first-order systems design constraint, not an accounting nuance.

\bibliographystyle{ACM-Reference-Format}
\bibliography{sample-base}
\newpage
\appendix

\section{Performance Model Details}
\label{sec:appendix-perf-model}

We use a first-order comparative model of MoE inference throughput. For a
given deployment, throughput is limited by the slowest of compute, HBM
bandwidth, and communication. The model is used only to compare hardware and
locality configurations in the fleet study, not to predict absolute serving
latency.

\subsection{Per-Phase Throughput}
\label{sec:appendix-throughput}

For model $m$, deployment $D$, and phase
$\phi \in \{\mathrm{pre}, \mathrm{dec}\}$, we write
\begin{equation}
    \mathrm{TPS}^{\phi}(m,D) =
    \min\!\left(
        \frac{F_D}{\mathcal{C}^{\phi}(m)},
        \frac{B_D^{\mathrm{HBM}}}{\mathcal{M}^{\phi}(m)},
        \frac{1}{T_{\mathrm{comm}}^{\phi}(m,D)}
    \right),
    \label{eq:phase-tps}
\end{equation}
where $F_D$ is deployment compute throughput, $B_D^{\mathrm{HBM}}$ is aggregate
HBM bandwidth, $\mathcal{C}^{\phi}$ and $\mathcal{M}^{\phi}$ are the per-token
compute and memory costs of phase $\phi$, and
$T_{\mathrm{comm}}^{\phi}$ is the communication time.

We assume $1$ FMA $= 2$ FLOPs. Unless otherwise noted, we use FP8 weights
($b_w = 1$ byte), FP4 activations and KV cache
($b_{\mathrm{act}} = b_{\mathrm{kv}} = 0.5$ bytes), and serving batch size
$B = 256$. All models use $K=2$ routed experts per token and
$\mathrm{FF}=4w$.

We use the same bottleneck form for prefill and decode; the phases differ only
in weight traffic, KV traffic, and sequence-length dependence:
\begin{align}
    \mathcal{C}^{\mathrm{pre}}(m)
    &= L\!\left(4K w\mathrm{FF} + 4w^2 + 2wS_p\right),\\
    \mathcal{C}^{\mathrm{dec}}(m,t)
    &= L\left(4K w\mathrm{FF} + 4w^2 + 2wt\right), \label{eq:c-phases} \\
    \mathcal{M}^{\mathrm{pre}}(m)
    &\approx \frac{W_{\mathrm{total}}(m)}{B S_p} + 2Lwb_{\mathrm{kv}}, \\
    \mathcal{M}^{\mathrm{dec}}(m,t)
    &\approx \frac{W_{\mathrm{active}}(m)}{B} + 2Lw(t+1) b_{\mathrm{kv}}, \label{eq:m-phases} \\
    \mathcal{N}_{\mathrm{TP}}^{\phi}(m,D)
    &= L \cdot \frac{2(T_D-1)}{T_D} w b_{\mathrm{act}}, \\
    \mathcal{N}_{\mathrm{EP}}^{\phi}(m)
    &= 2LKwb_{\mathrm{act}}. \label{eq:n-phases}
\end{align}

Here $L$ is the number of transformer layers, $w$ is hidden width, $S_p$ is
prompt length, and $t$ is the effective context length during decode.
$W_{\mathrm{total}}$ counts all expert parameters, while
$W_{\mathrm{active}}$ counts the shared attention weights plus the routed
experts touched by one token.

Within one NVLink domain, shared attention uses tensor parallelism across
$T_D$ packages, while MoE FFNs use expert parallelism. TP traffic stays on
NVLink. EP traffic is split between local NVLink and remote InfiniBand
according to the locality model below. Prefill amortizes total weight traffic
across the prompt batch, whereas decode depends on active weights and growing
KV-cache reads.

\subsection{Communication Locality Model}
\label{sec:appendix-capacity}

Communication depends on how much expert-parallel traffic remains within one
local high-bandwidth domain. We estimate this from model fit in HBM.

Let $N_{\mathrm{pkg}}^D$ be the number of accelerator packages in one local
domain of deployment $D$, and let $\mathrm{HBM}_{\mathrm{pkg}}^D$ be HBM
capacity per package. We reserve a fraction $1-\alpha$ of HBM for KV
residency and runtime overhead, with $\alpha = 0.7$. The number of local
NVLink domains required to host model $m$ on deployment $D$ is
\begin{equation}
    N_{\mathrm{dom}}(m,D) =
    \left\lceil
        \frac{W_{\mathrm{total}}(m)}
             {\alpha\,N_{\mathrm{pkg}}^{D}\,\mathrm{HBM}_{\mathrm{pkg}}^{D}}
    \right\rceil.
    \label{eq:n-dom}
\end{equation}

We then approximate the fraction of expert-parallel traffic that leaves the
local NVLink domain as
\begin{equation}
    f_{\mathrm{IB}}(m,D)=
    \begin{cases}
        0, & N_{\mathrm{dom}}(m,D)=1, \\[3pt]
        1-\dfrac{1}{N_{\mathrm{dom}}(m,D)}, & \text{otherwise}.
    \end{cases}
    \label{eq:fib-full}
\end{equation}
This captures the first-order fact that once a model spans multiple domains,
only a fraction $1/N_{\mathrm{dom}}$ of expert traffic can remain local.

We use package counts throughout. HBM capacity, package TDP, and
NVLink-domain membership are package-level quantities, so all capacity and
communication terms are evaluated at the package/domain level.

Given $f_{\mathrm{IB}}(m,D)$, we model TP and EP communication time as

\begin{align}
    T_{\mathrm{TP}}^\phi(m,D)
    &= \frac{\mathcal{N}_{\mathrm{TP}}^\phi(m,D)}{B_D^{\mathrm{NVL}}}, \label{eq:t-tp} \\[2pt]
    T_{\mathrm{EP}}^\phi(m,D)
    &= \max\!\left(
        \frac{(1-f_{\mathrm{IB}}(m,D))\mathcal{N}_{\mathrm{EP}}^\phi(m)}
             {B_D^{\mathrm{NVL}}},
        \frac{f_{\mathrm{IB}}(m,D)\mathcal{N}_{\mathrm{EP}}^\phi(m)}
             {B_D^{\mathrm{IB}}}
    \right), \label{eq:t-ep} \\[2pt]
    T_{\mathrm{comm}}^\phi(m,D)
    &= T_{\mathrm{TP}}^\phi(m,D) + T_{\mathrm{EP}}^\phi(m,D), \label{eq:t-comm}
\end{align}
where $B_D^{\mathrm{NVL}}$ and $B_D^{\mathrm{IB}}$ are local NVLink and remote
InfiniBand bandwidth. The $\max$ term reflects concurrent local and remote
transfers during the EP sublayer.

\subsection{Request-Level Throughput}
\label{sec:appendix-request-throughput}

For prompt length $S_p$ and output length $S_{\mathrm{out}}$, we aggregate
prefill, decode, and disaggregated KV transfer into a request-level
throughput:
\begin{equation}
    \mathrm{TPS}(m,D)
    =
    \frac{B S_{\mathrm{out}}}{
        \dfrac{B S_p}{\mathrm{TPS}^{\mathrm{pre}}(m,D)}
        +
        \displaystyle\sum_{t=S_p+1}^{S_p+S_{\mathrm{out}}}
        \frac{1}{\mathrm{TPS}^{\mathrm{dec}}(m,t,D)}
        +
        T_{\mathrm{KV}}
    }.
    \label{eq:combined-tps-appendix}
\end{equation}
Here $\mathrm{TPS}^{\mathrm{pre}}(m,D)$ is evaluated at prompt length $S_p$, and
\begin{equation}
    T_{\mathrm{KV}} = \frac{2Lw\,S_p\,b_{\mathrm{kv}}}{B_{\mathrm{transfer}}}
    \label{eq:kv-transfer}
\end{equation}
captures KV transfer for disaggregated serving~\cite{distserve,splitwise}.

\subsection{Model Inputs and Limitations}
\label{sec:appendix-limitations}

The model consumes the following model-specific inputs:
$(L, w, W_{\mathrm{total}}, W_{\mathrm{active}})$.
For the model suite used in the paper, all models use $K=2$ routed experts per
token and $\mathrm{FF}=4w$.

This model has three deliberate limitations.

\begin{enumerate}
    \item It is a first-order comparative model, not a topology-accurate
    runtime simulator.
    \item Communication is modeled with bandwidth-time approximations rather
    than collective-specific kernels.
    \item We do not model fine-grained overlap among TP communication, EP
    communication, and compute.
\end{enumerate}


\subsection{Workload Model Suite}
\label{sec:model-suite}

Table~\ref{tab:model-suite} lists the model configurations used in the throughput study.
The MoE suite spans three orders of magnitude in total parameters,
from a 0.6\,T model whose experts fit within a single rack-local
NVLink domain to a 401\,T model that requires expert-parallel
communication across multiple domains.
All MoE models use top-$K{=}2$ routing and $\text{FF}{=}4w$.
Expert counts grow with model size, which increases the fraction of
traffic that spills onto inter-domain links for a given deployment
architecture.

\begin{table}[t]
\centering
\caption{Model configurations for the inference throughput study.
$L$: transformer layers; $w$: hidden dimension; $E$: total experts;
$K$: routed experts per token; $S$: evaluation context length.
All MoE models use $\text{FF}=4w$.}
\label{tab:model-suite}
\small
\begin{tabular}{lrrrrr}
\toprule
Model & $L$ & $w$ & $E$ & $K$ & $S$ \\
\midrule
MoE-0.6T   &  48 &  6\,144 &    64 & 2 & 1\,024 \\
MoE-5T     &  96 &  8\,192 &    96 & 2 & 1\,024 \\
MoE-19T    & 120 & 12\,288 &   128 & 2 & 1\,024 \\
MoE-51T    & 120 & 14\,336 &   256 & 2 & 1\,024 \\
MoE-132T   & 120 & 16\,384 &   512 & 2 & 1\,024 \\
MoE-401T   & 144 & 18\,432 & 1\,024 & 2 & 1\,024 \\
\bottomrule
\end{tabular}
\end{table}

\section{Hardware and Cost Projections}
\label{sec:appendix-projections}

We instantiate the fleet study with comparative projections for GPU package
power, GPU package capability, deployment-unit architecture, non-GPU rack
power, and facility infrastructure cost. Package-level trends determine
compute, HBM bandwidth, HBM capacity, and TDP growth. Deployment architecture
determines rack power, local-domain size, and communication bandwidth. These
assumptions are used for sensitivity analysis, not product forecasting.

\subsection{GPU Deployment Projections}
\label{sec:appendix-gpu-projections}

A GPU \emph{package} is the atomic unit of the projection model. A package may
contain one or more compute dies together with its attached HBM. Package-level
quantities determine TDP, compute throughput, HBM bandwidth, and HBM capacity.

We vary package TDP across Low, Medium, and High scenarios:
\begin{equation}
    P_{\mathrm{pkg}}(\tau,s)
    =
    P_{\mathrm{pkg}}^{\mathrm{anchor}}(s)\,(1+g_s)^{\tau-\tau_{\mathrm{anchor}}},
    \label{eq:tdp-growth}
\end{equation}
where \(s\in\{\mathrm{Low},\mathrm{Med},\mathrm{High}\}\) and
\(g_s\in\{5\%,\,12.5\%,\,20\%\}\).

We hold performance to central projections so that scenario variation reflects
power-density uncertainty rather than simultaneous changes in compute
capability. For publicly disclosed near-term hardware, we use vendor-reported
package-level anchors. For later years, we extrapolate FP4 FLOP/s, HBM
bandwidth, and HBM capacity at constant annual rates of \(30\%\), \(15\%\),
and \(25\%\), respectively. Oberon is anchored at B200 in 2025 and Vera Rubin
in 2026. Our later pod-scale study case is anchored at Rubin Ultra in 2027,
held fixed through 2028, and extrapolated beginning in 2029.

Deployment architecture specifies how packages are integrated into one
deployment unit: package count, local NVLink-domain size, aggregate NVLink
bandwidth, aggregate scale-out bandwidth, and non-package overhead power.
These quantities change only at architectural transitions.

\begin{table*}[t]
\centering
\caption{Deployment architecture parameters. NVLink values are aggregate
unidirectional bandwidth per local NVLink domain; scale-out values are
aggregate per deployment unit.}
\label{tab:arch-params}
\small
\begin{tabular}{llcccccc}
\toprule
& & \(N_{\mathrm{pkg}}\) & Dies & NVL domain & \(B^{\mathrm{NVL}}_D\) & \(B^{\mathrm{IB}}_D\) & \(P_{\mathrm{ovhd}}\) \\
Architecture & Available & (pkgs) & /pkg & (pkgs) & (TB/s) & (TB/s) & (kW) \\
\midrule
DGX-H200              & 2024   &   8 & 1 &   8 &   3.6 &  0.4 &  3 \\
Blackwell--Oberon     & 2025   &  72 & 1 &  72 &  64.8 &  7.2 & 25 \\
Vera Rubin NVL72      & 2026+  &  72 & 2 &  72 & 259.2 & 14.4 & 30 \\
Kyber / Rubin Ultra   & 2027+  & 144 & 4 & 144 & 750.0 & 57.6$^\dagger$ & 35 \\
\bottomrule
\end{tabular}
\vspace{2pt}

\raggedright\footnotesize
$^\dagger$Later pod-scale study assumption derived from public Rubin Ultra
system disclosures under our unidirectional aggregate convention; detailed NIC
topology remains unsettled in public materials.
\end{table*}

Given package projections and deployment architecture parameters, we derive the
rack-level quantities consumed by the throughput and placement models:
\begin{align}
    F_D(\tau) &= N_{\mathrm{pkg}} \cdot F_{\mathrm{pkg}}(\tau), \label{eq:derived-flops} \\
    B_D^{\mathrm{HBM}}(\tau) &= N_{\mathrm{pkg}} \cdot B_{\mathrm{pkg}}^{\mathrm{HBM}}(\tau), \label{eq:derived-hbm-bw} \\
    H_D^{\mathrm{usable}}(\tau) &= \alpha \cdot N_{\mathrm{pkg}} \cdot \mathrm{HBM}_{\mathrm{pkg}}(\tau), \label{eq:derived-hbm-cap} \\
    P_{\mathrm{rack}}(\tau,s) &= N_{\mathrm{pkg}} \cdot P_{\mathrm{pkg}}(\tau,s) + P_{\mathrm{ovhd}}. \label{eq:rack-power}
\end{align}
Here \(B_D^{\mathrm{NVL}}\) and \(B_D^{\mathrm{IB}}\) are taken directly from
Table~\ref{tab:arch-params}. We use package counts throughout, since HBM
capacity, package TDP, and NVLink-domain membership are package-level
quantities.

\begin{table*}[t]
\centering
\caption{Per-package performance projections used to instantiate the throughput
model. Oberon is anchored at B200 (2025) and Vera Rubin (2026); the later
pod-scale study case is anchored at Rubin Ultra (2027). Post-anchor
extrapolation begins in 2029.}
\label{tab:pkg-perf-projections}
\small
\begin{tabular}{l ccc ccc}
\toprule
& \multicolumn{3}{c}{Oberon} & \multicolumn{3}{c}{Kyber / Rubin Ultra} \\
\cmidrule(lr){2-4} \cmidrule(lr){5-7}
Year & \(F\) (PF) & \(B^{\mathrm{HBM}}\) (TB/s) & HBM (GB) & \(F\) (PF) & \(B^{\mathrm{HBM}}\) (TB/s) & HBM (GB) \\
\midrule
2025 & 10.0 & 8.0 & 192 &  &  &  \\
2026 & 50.0 & 22.0 & 288 &  &  &  \\
2027 & 50.0 & 22.0 & 288 & 100.0 & 32.0 & 1{,}024 \\
2028 & 50.0 & 22.0 & 288 & 100.0 & 32.0 & 1{,}024 \\
2029 & 65.0 & 25.3 & 360 & 130.0 & 36.8 & 1{,}280 \\
2030 & 84.5 & 29.1 & 450 & 169.0 & 42.3 & 1{,}600 \\
2031 & 109.9 & 33.5 & 563 & 219.7 & 48.7 & 2{,}000 \\
2032 & 142.8 & 38.5 & 703 & 285.6 & 56.0 & 2{,}500 \\
2033 & 185.6 & 44.2 & 879 & 371.3 & 64.4 & 3{,}125 \\
2034 & 241.3 & 50.9 & 1{,}099 & 482.7 & 74.0 & 3{,}906 \\
\bottomrule
\end{tabular}
\end{table*}

\begin{table*}[t]
\centering
\caption{Derived rack power (kW) across growth scenarios. Anchor values follow
announced or study-anchor specifications; later values are extrapolated using
Eq.~\ref{eq:rack-power}.}
\label{tab:unified-rack-power}
\small
\begin{tabular}{c ccc c ccc}
\toprule
& \multicolumn{3}{c}{Oberon (\(N_{\mathrm{pkg}}=72\))} & & \multicolumn{3}{c}{Kyber / Rubin Ultra (\(N_{\mathrm{pkg}}=144\))} \\
\cmidrule(lr){2-4} \cmidrule(lr){6-8}
Year & Low & Med & High & & Low & Med & High \\
\midrule
2025 & 157 & 180 & 203 & & --- & --- & --- \\
2026 & 160 & 178 & 196 & & --- & --- & --- \\
2027 & 166 & 197 & 226 & & 515 & 600 & 685 \\
2028 & 173 & 218 & 262 & & 515 & 600 & 685 \\
\cmidrule(lr){1-8}
2029 & 180 & 243 & 341 & & 539 & 671 & 815 \\
2030 & 188 & 271 & 434 & & 564 & 750 & 971 \\
2031 & 197 & 303 & 545 & & 591 & 839 & 1{,}158 \\
2032 & 205 & 339 & 677 & & 619 & 940 & 1{,}382 \\
2033 & 214 & 379 & 836 & & 648 & 1{,}053 & 1{,}652 \\
2034 & 224 & 425 & 1{,}025 & & 679 & 1{,}180 & 1{,}975 \\
\bottomrule
\end{tabular}
\end{table*}

\subsection{Pod and Non-GPU Assumptions}
\label{sec:appendix-pod-nongpu}

A deployment pod is a co-procured, co-placed multi-rack unit. We distinguish
between \emph{rack-scale deployment units} and \emph{pod-scale deployment
units}. In the baseline model, each rack retains its own local NVLink domain,
so pods change placement quantum but not rack-local communication structure:
\begin{equation}
    N_{\mathrm{NVL}}^{\mathrm{pod}} = N_{\mathrm{NVL}}^{\mathrm{rack}}.
    \label{eq:pod-nvl}
\end{equation}
Pod power is the sum of the constituent racks:
\begin{equation}
    P_{\mathrm{pod}}(\tau,s) = \sum_{r \in \mathcal{R}_{\mathrm{pod}}} P_r(\tau,s),
    \label{eq:pod-power}
\end{equation}
where \(\mathcal{R}_{\mathrm{pod}}\) is the set of rack types in the pod. For a
homogeneous pod, this reduces to \(N_{\mathrm{racks}} P_{\mathrm{rack}}\).

General-compute racks are anchored at 20\,kW in 2025 and grow at
\(\{3\%,\,5\%,\,8\%\}\) annually, reaching \(\{26,\,38,\,52\}\)\,kW by 2034.
Storage racks are anchored at 15\,kW in 2025 and grow at
\(\{2\%,\,4\%,\,6\%\}\) annually, reaching \(\{18,\,22,\,26\}\)\,kW by 2034.
These trajectories define the non-GPU rack power inputs used by the SKU
generation procedure. Unless otherwise noted, compute and storage use the
Medium scenario and GPU racks vary across scenarios.

\subsection{Facility Cost Assumptions}
\label{sec:appendix-cost}

The cost model is used only to compare designs under a common infrastructure
baseline and to instantiate the initial and effective \$/MW metrics in the main
evaluation. It is not intended to predict operator-specific build cost. Values
include equipment and installation and are intended as representative installed
costs rather than hyperscaler-specific internal costs.

\begin{table*}[t]
\centering
\small
\caption{Facility infrastructure cost assumptions per MW of IT capacity.}
\label{tab:component_costs}
\begin{tabular}{lr}
\toprule
Component & Cost/MW \\
\midrule
UPS systems & \$1{,}000{,}000 \\
Battery systems & \$275{,}000 \\
Backup generators & \$750{,}000 \\
MV transformers & \$120{,}000 \\
MV switchgear & \$60{,}000 \\
LV switchboards & \$150{,}000 \\
Automatic transfer switches & \$70{,}000 \\
Static transfer switches & \$250{,}000 \\
Row distribution (PDUs/busway) & \$100{,}000 \\
Busbar overhead & \$6{,}000 \\
Cooling systems & \$3{,}000{,}000 \\
Facility shell, site \& engineering & \$1{,}800{,}000 \\
Fit-out \& other & \$2{,}800{,}000 \\
\bottomrule
\end{tabular}
\end{table*}

\section{Datacenter Reference Designs, Placement, and Parameters}

\subsection{Placement Feasibility}
\label{app:placement-formalization}

This subsection formalizes the hierarchical placement constraint used by the
simulator.

\paragraph{Demand vector.}
Each deployment unit $r$ has resource demand vector
\[
\mathbf{d}_r = \left(P_r, \mathrm{CFM}_r, \mathrm{LPM}_r, n_r\right),
\]
where $P_r$ is power demand (kW), $\mathrm{CFM}_r$ is air-cooling demand,
$\mathrm{LPM}_r$ is liquid-cooling demand, and $n_r$ is tile count. GPU pods
may consume all four resources. General-compute and storage racks have
$\mathrm{LPM}_r = 0$.

\paragraph{Ancestor-path feasibility.}
We model the power-delivery hierarchy as a rooted tree whose internal nodes are
distribution components with per-resource capacities. For a candidate row
location $\ell$, let
\[
\mathrm{path}(\ell) = \{\ell_0, \ell_1, \dots, \ell_h\}
\]
denote the ancestor path from the row ($\ell_0$) to the substation
($\ell_h$). Placement of deployment $r$ at location $\ell$ is feasible if and
only if
\begin{equation}
L^{(m)}_{\ell_k} + d_r^{(m)} \le C^{(m)}_{\ell_k,\mathrm{eff}}
\qquad
\forall \ell_k \in \mathrm{path}(\ell),\ \forall m,
\label{eq:feasibility}
\end{equation}
where $L^{(m)}_{\ell_k}$ is the current aggregate load at node $\ell_k$ in
resource dimension $m$, $d_r^{(m)}$ is the corresponding component of
$\mathbf{d}_r$, and $C^{(m)}_{\ell_k,\mathrm{eff}}$ is the effective capacity
at that node after accounting for redundancy constraints.

\paragraph{Effective capacity under redundancy.}
Effective electrical capacity depends on redundancy topology and availability
tier.

For distributed redundancy $(xN/y)$, a system with $x$ total line-ups and $y$
line-ups of usable high-availability capacity reserves failover headroom within
each active line-up. For a high-availability deployment,
\begin{equation}
C^{\mathrm{eff}}_{\ell} = \frac{y}{x}\, C_{\ell},
\label{eq:dist-eff}
\end{equation}
at each line-up node $\ell$. Low-availability deployments may use the full
rated capacity $C_{\ell}$ and therefore consume reserve capacity.

For block redundancy $(N+k)$, $N$ primary line-ups carry IT load and $k$ standby line-ups are reserved for failover. Each primary line-up may therefore be loaded to its rated capacity, so $C^{\mathrm{eff}}_{\ell} = C_{\ell}$ for all placed deployments. The standby line-ups do not constrain placement at the row level, but they do contribute to total hall cost.

\paragraph{Non-power resources.}
Cooling and space constraints are enforced at row and ancestor nodes exactly as
in Eq.~\ref{eq:feasibility}. Networking is provisioned at build time and is not
modeled as a binding online placement constraint. Space is limited by the fixed
tile count per row.

\subsection{Counting Rows}
\label{sec:appendix-rows}

Reference designs are defined by a set of UPS line-ups, a partition into power
domains, and balanced row-to-line-up wiring. Balance means that all distinct
connection patterns allowed by a design appear equally often.

\paragraph{Block-redundant designs.}
In a block-redundant design, all rows in one power domain connect to the same
set of active line-ups. If a hall has $N$ line-ups partitioned into $k$ power
domains, then the row count in each class must be a multiple of $\frac{N}{k}.$

\paragraph{Distributed-redundant designs.}
In a distributed-redundant design, balance requires all admissible connection
combinations within a power domain to be represented equally. We use two row
classes. Low-density rows connect to two upstream line-ups, so their count must
be a multiple of
$\binom{N/k}{2}.$
High-density rows connect to four upstream line-ups, so their count must be a
multiple of
$\binom{N/k}{4}.$

\paragraph{Row classes.}
Each row contains 24 rack positions. Low-density rows use two upstream feeds.
High-density rows use four. This two-class model is a stylized approximation
used to compare designs under common assumptions; it does not exclude other
feed counts or row constructions.

For block-redundant designs, we use a base hall with $6N$ low-density rows and
$4N$ high-density rows. For distributed-redundant designs, row counts must
respect the balanced-combination rules above. We therefore choose the smallest
integer multiples of $\binom{N/k}{2}$ and $\binom{N/k}{4}$ that make the ratio
of high-density to low-density rows as close as possible to the block-design
reference while keeping total row counts comparable.

We bias the reference designs toward more low-density rows because low-density rows absorb residual line-up capacity that cannot be consumed by high-power deployments once high-density rows saturate. Without enough low-density rows, capacity strands at the line-up level even when aggregate hall power remains available.

\end{document}